\documentclass[numberedappendix]{emulateapj}
\usepackage{natbib}
\usepackage{graphicx}

\slugcomment{Accepted for publication in ApJ on 10 October 2010}

\def\sun{\ifmmode\odot\else$\odot$\fi}
\def\micron{\hbox{\,$\mu$m}}
\def\estrella{$\star$}

\newcommand{\Neii}{\hbox{[\ion{Ne}{2}]12.81\micron}}
\newcommand{\Neiii}{\hbox{[\ion{Ne}{3}]15.56\micron}}
\newcommand{\Neva}{\hbox{[\ion{Ne}{5}]14.32\micron}}
\newcommand{\Nevb}{\hbox{[\ion{Ne}{5}]24.32\micron}}
\newcommand{\Oiv}{\hbox{[\ion{O}{4}]25.89\micron}}
\newcommand{\HII}{H\,{\sc ii}}
\newcommand{\kboltz}{\textit{k}}

\def\Spitzer{\textit{Spitzer}}

\shorttitle{}
\shortauthors{Pereira-Santaella et al.}

\begin{document}

\title{The Mid-Infrared high-ionization lines from Active Galactic Nuclei and star forming galaxies$^{\star}$}
\author{Miguel Pereira-Santaella\altaffilmark{1,2}}
\email{pereira@damir.iem.csic.es}
\author{Aleksandar~M. Diamond-Stanic\altaffilmark{3}}
\author{Almudena Alonso-Herrero\altaffilmark{1,2,4}}
\author{George~H. Rieke\altaffilmark{3}}

\altaffiltext{$\star$}{This work is based on observations made with the Spitzer Space Telescope, which is operated by the Jet Propulsion Laboratory, California Institute of Technology under NASA contract 1407}
\altaffiltext{1}{Instituto de Estructura de la Materia, CSIC, Serrano 121, E-28006, Madrid, Spain}
\altaffiltext{2}{Departamento de Astrof\'isica, Centro de Astrobiolog\'ia, CSIC/INTA, Carretera de Torrej\'on a Ajalvir, km 4, 28850, Torrej\'on de Ardoz, Madrid, Spain}
\altaffiltext{3}{Steward Observatory, University of Arizona, 933 North Cherry Avenue, Tucson, AZ 85721, USA}
\altaffiltext{4}{Associate Astronomer, Steward Observatory, University of Arizona, AZ 85721, USA}

\begin{abstract}
We used \Spitzer/IRS spectroscopic data on 426 galaxies including quasars, Seyferts, LINER and \HII\ galaxies to investigate the relationship among the mid-IR emission lines.
There is a tight linear correlation between the [\ion{Ne}{5}]14.3\micron\ and 24.3\micron\ (97.1\,eV) and the [\ion{O}{4}]25.9\micron\ (54.9\,eV) high-ionization emission lines. The correlation also holds for these high-ionization emission lines and the \Neiii\ (41\,eV) emission line, although only for active galaxies.
We used these correlations to calculate the [\ion{Ne}{3}] excess due to star formation in Seyfert galaxies.
We also estimated the [\ion{O}{4}] luminosity due to star formation in active galaxies and determined that it dominates the [\ion{O}{4}] emission only if the contribution of the active nucleus to the total luminosity is below 5\%. We find that the AGN dominates the [\ion{O}{4}] emission in most Seyfert galaxies, whereas star-formation adequately explains the observed [\ion{O}{4}] emission in optically classified \HII\ galaxies.
Finally we computed photoionization models to determine the physical conditions of the narrow line region where these high-ionization lines originate. The estimated ionization parameter range is -2.8 $<$ $\log$ $U$ $<$ -2.5 and the total hydrogen column density range is 20 $<$ $\log$ $n_{\rm H}$ (cm$^{-2}$) $<$ 21.
\end{abstract}


\section{Introduction}\label{s:intro}

Primarily through measurements in the optical, active galactic nuclei (AGNs) have been categorized as Type 1 or 2, respectively with and without very broad emission lines. The unification model \citep{Antonucci1993, Urry1995} successfully explains this behavior in terms of a dusty circumnuclear torus that hides the broad-line region (BLR) from our line of sight for type 2 AGN. 
It also appears that extinction in the host galaxy can hide the BLR (e.g., \citealt{Maiolino1995, AAH2003}) in some active galaxies.
Extinction therefore affects the optical properties of AGN in fundamental ways. These effects can be minimized by studying these objects using infrared (IR) emission lines, both to test the predictions of the unification model and to characterize AGNs in a uniform way. 
Toward this end, we have used mid-IR observations with the Infrared Spectrograph \citep[IRS,][]{HouckIRS} on \Spitzer\ to examine the behavior of a large sample of AGN. 

Important high-ionization lines accessible in the mid-IR include [\ion{Ne}{5}] (97.1\,eV) at 14.32 and 24.32\micron, [\ion{O}{4}] (54.9\,eV) at 25.89\micron, and [\ion{Ne}{3}] (41\,eV) at 15.56\micron.
Because of their very high ionization potential, the [\ion{Ne}{5}] lines are considered to be reliable signposts for an AGN \citep{Genzel1998,Armus07}. These lines have been used to estimate the accretion power in the local Universe \citep{Tommasin2010} and to identify low-luminosity AGNs in local galaxies \citep{Satyapal2008, Goulding2009}. 
However, they are also produced in supernova remnants \citep[SNR,][]{Oliva1999, Smith2009}, planetary nebulae \citep[PN,][]{Pottasch2009}, and Wolf-Rayet stars \citep[WR,][]{Schaerer1999}. Although the [\ion{Ne}{5}] luminosities of these objects are low, $\sim 10^{34}$\,erg s$^{-1}$ \citep{Smith2009, Pottasch2009}, several thousands of them might also produce detectable [\ion{Ne}{5}] emission.

Despite the lower ionization potential of [\ion{O}{4}], the 25.89\micron\ emission line appears to be an accurate indicator of AGN power, since it correlates well with the hard ($>$ 14\,keV) X-ray luminosity \citep{Melendez2008, Rigby2009, Diamond2009} and the mid-IR [\ion{Ne}{5}] emission lines \citep{Dudik2009, Weaver2010}. However, this line also appears in the spectra of starburst galaxies with no other evidence for AGN \citep{Lutz1998, Bernard-Salas2009}, where it is attributed to Wolf-Rayet stars \citep{Crowther1999, Schaerer1999} and/or to shocks \citep{Allen2008, Lutz1998}. 

The [\ion{Ne}{3}] 15.56\micron\ line is excited by young, massive stars \citep{Verma03,Brandl06,Beirao2006,Beirao08,Ho07,Bernard-Salas2009,AAH09Arp299,Pereira2010}. Nonetheless, it seems to be a reasonably good proxy for AGN luminosity, at least for reasonably high luminosities \citep{Gorjian2007, Dudik2009, Melendez2008, Tommasin08}. 
Quantifying when its excitation is dominated by an active nucleus would allow it to be used in concert with the other high-ionization lines to probe conditions in the neighborhoods of AGNs.  

In this paper, we present a statistical study of the behavior of these lines in a sample of 426 galaxies. The data are compiled from the literature on high spectral resolution ($R\,\sim$\,600) IRS measurements, as well as our own new measurements of low-luminosity Seyfert galaxies and luminous infrared galaxies (LIRGs).  We describe the sample and the data reduction respectively in Sections 2 and 3. 
Section 4 explores the correlations among the high-ionization emission lines. In Section 5, we  discuss the star formation contribution to these lines. Section 6 compares the star formation in Seyfert 1 and 2 galaxies. Finally, in Section 7 we use photoionization models to study the physical conditions in the narrow line regions (NLRs).

Throughout this paper we assume a flat cosmology with $H_0 = 70$ km s$^{-1}$Mpc$^{-1}$, $\Omega_{\rm M} = 0.3$, and $\Omega_{\rm \Lambda} = 0.7$.

\section{The Sample}\label{s:sample}

The sample contains 426 galaxies (Table \ref{tbl_fluxes_sample}) for which high spectral resolution ($R\,\sim$\,600) \Spitzer\slash IRS spectra were available, either in the literature (\citealt{Weedman2005, Ogle2006, Farrah07, Gorjian2007, Tommasin08, Tommasin2010, Veilleux2009, Dale2009, Bernard-Salas2009, Goulding2009}; Pereira-Santaella et al. 2011, in prep.), in the \Spitzer\ archive or observed through the programs 40936 and 50597 (PI: G. H. Rieke).
AGN type classifications were taken from NASA Extragalactic Database (NED). We designated only pure cases as Type 2 and included within Type 1 all intermediate cases (i.e.: 1.2, 1.5, 1.8, 1.9) based on observations and modeling of the nuclear spectra energy distributions \citep{AAH2003, RamosAlmeida2009}.
We could not find the nuclear activity classification for 32 of the galaxies.
The sample includes 28 QSOs, 76 Seyfert 1, 125 Seyfert 2, 55 LINERs and 110 \HII\ or starburst galaxies (see Table \ref{tbl_sample}).
The luminosity of the active galaxies ranges from QSO (QUEST sample, \citealt{Veilleux2009}) to typical Seyfert galaxies (12\micron\ sample, \citealt{Rush1993, Tommasin2010}), relatively low-luminosity Seyferts (RSA sample, \citealt{Maiolino1995, Ho1997, Diamond2009}) and LINERs \citep{Sturm2006}.
According to their infrared luminosities 71 sample members are classified as LIRGs ($L_{\rm IR} = 10^{11}-10^{12} L_{\rm \odot}$) and 54 as ultraluminous infrared galaxies (ULIRGs, $L_{\rm IR} > 10^{12} L_{\rm \odot}$).
We found in the Revised Bright Galaxy Sample catalog \citep{SandersRBGS, Surace2004} the \textit{IRAS} fluxes for 196 of our galaxies which we used to calculate their $L_{\rm IR}$ as defined in \citet{Sanders96}.

Note that the LINER group includes very different galaxies. IR-bright LINERs ($L_{\rm IR}$/$L_{\rm B}$ $\gtrsim$ 1) have infrared SEDs similar to starbursts, although high-ionization lines and hard X-ray cores are detected in some, suggesting the presence of an AGN. In comparison, IR-faint LINERs ($L_{\rm IR}$/$L_{\rm B}$ $\lesssim$ 1) seem to be powered by an AGN \citep{Sturm2006, Satyapal2004}. 

\begin{figure}[h]
\center
\includegraphics[width=0.45\textwidth]{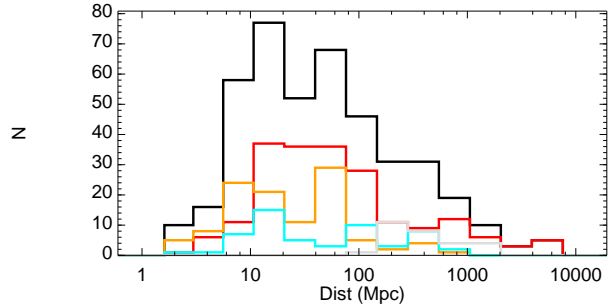}
\caption{Distribution of the galaxy distances of the complete sample (black), QSO (gray), Seyfert 1 and 2 galaxies (red), LINERs (blue) and \HII\ galaxies (orange).}
\label{fig_distance}
\end{figure}

Figure \ref{fig_distance} shows the distance distribution of the sample. The median distance is 60 Mpc and most (68\%) of the galaxies are between 15 and 400 Mpc.

\setcounter{table}{1}
\tabletypesize{\small}
\begin{deluxetable}{lccccl}
\tablewidth{0pt}
\tablecaption{The sample\label{tbl_sample}}
\tablehead{ \colhead{Type} & \colhead{N} & \colhead{[\ion{Ne}{5}]14.3\micron} & \colhead{[\ion{Ne}{5}]24.3\micron} & \colhead{[\ion{O}{4}]}}
\startdata
QSO & 28 & 22 & 16 & 25\\
Seyfert 1 & 76 & 59 & 42 & 63\\
Seyfert 2 & 125 & 89 & 69 & 101\\
LINER & 55 & 10 & 6 & 27\\
\HII\slash Starburst & 110 & 4 & 2 & 59\\
Unknown & 32 & 2 & 1 & 7\\
Total & 426 & 186 & 136 & 282
\enddata
\tablecomments{N is the number of galaxies of each type. For each type we give the number of detections of the \Neva, \Nevb\ and \Oiv\ emission lines.}
\end{deluxetable}

\tabletypesize{\small}
\begin{deluxetable*}{ccccccccc}
\tablewidth{0pt}
\tablecaption{Median line ratios\label{tbl_ratios}}
\tablehead{  & \multicolumn{5}{c}{Median ratio}\\
 \cline{2-6}  \\[-1.5ex]
\colhead{Ratio} &  \colhead{QSO} & \colhead{Sy1} & \colhead{Sy2}& \colhead{LINER}& \colhead{\HII}}
\startdata
\Oiv\slash\Nevb   & 3.3$\pm$1.2\tablenotemark{\estrella} & 3.6$\pm$0.6 & 3.5$\pm$0.6 & 3.7$\pm$1.0\tablenotemark{\estrella} & 4.1$\pm$1.3\tablenotemark{\estrella} \\
\Oiv\slash\Neva   & 3.6$\pm$0.9 & 3.5$\pm$1.0 & 3.5$\pm$1.3 & 4.8$\pm$2.0\tablenotemark{\estrella} & 4.4$\pm$2.0\tablenotemark{\estrella} \\
\Nevb\slash\Neva  & 1.0$\pm$0.3\tablenotemark{\estrella} & 1.1$\pm$0.3 & 1.0$\pm$0.3 & 1.3$\pm$0.3\tablenotemark{\estrella} & 4.7$\pm$4.1\tablenotemark{\estrella} \\
\Neiii\slash\Neva & 1.7$\pm$0.5 & 1.9$\pm$0.6 & 2.0$\pm$0.7 & 5.3$\pm$3.7\tablenotemark{\estrella} & 6.8$\pm$1.7\tablenotemark{\estrella} \\
\Neii\slash\Neva  & 0.7$\pm$0.4 & 1.6$\pm$0.7 & 2.0$\pm$1.1 & 28$\pm$14\tablenotemark{\estrella}  & 57$\pm$42\tablenotemark{\estrella}  \\
\Neiii\slash\Oiv  & 0.5$\pm$0.2 & 0.6$\pm$0.2 & 0.9$\pm$0.4 & 2.3$\pm$1.1 & 4.2$\pm$2.4 \\
\Neii\slash\Oiv   & 0.2$\pm$0.1 & 0.5$\pm$0.3 & 1.1$\pm$0.9 & 6$\pm$2 & 22$\pm$16 \\
\Neiii\slash\Neii   & 2.0$\pm$1.0 & 1.2$\pm$0.6 & 0.8$\pm$0.4 & 0.3$\pm$0.2 & 0.17$\pm$0.07
\enddata
\tablecomments{Median ratios and uncertainties for each type of galaxies. The uncertainty is calculated as the median absolute deviation (should be multiplied by 1.48 to obtain the standard deviation).}
\tablenotetext{\estrella}{These values are calculated with less than 20 galaxies.}
\end{deluxetable*}

\section{Data reduction}\label{s:observations}

In addition to results from the literature, we utilized the measurements for the 88 galaxies of the RSA sample observed with the high spectral resolution modules
as well as for the 34 LIRGs of the \citet{AAH06s} sample (3 of these LIRGs are also members of the RSA sample).
We retrieved the basic calibrated data (BCD) from the \Spitzer\ archive processed by the pipeline version S18.7. We subtracted the background contribution when a dedicated sky observation was available. Note however that the sky subtraction is not important to measure these fine structure emission lines and it just improves the final quality of the spectra by removing bad pixels. Then we extracted the spectra using the standard programs included in the \Spitzer\ IRS Custom Extraction (SPICE) package provided by the \Spitzer\ Science Center (SSC). We assumed the point source calibration for all the galaxies, which is a good approximation for most of the galaxies at least for the high ionization lines.
For the galaxies observed in the mapping mode we extracted the nuclear spectra from the data cubes using a square aperture of 13\farcs4$\times$13\farcs4 and then we applied an aperture correction (Pereira-Santaella et al. 2011, in prep).
We used a Gaussian profile to fit the emission lines. For 9 galaxies (NGC~777, NGC~3254, NGC~3486, NGC~3941, NGC~4138, NGC~4378, NGC~4472, NGC~4698, NGC~5631) we were not able to measure any spectral feature due to the low signal-to-noise ratio of the spectra. They are not listed in Table \ref{tbl_fluxes_sample} nor included in the analyzed sample.

Table \ref{tbl_sample} includes the number of detections of the \Neva, \Nevb\ and \Oiv\ emission lines and in Table \ref{tbl_ratios} we show the observed median line ratios for each galaxy type both for galaxies from literature and those analyzed by us. Note that there are 4 detections of the \Neva\ line and 2 of the \Nevb\ line in galaxies classified as \HII. For these galaxies (NGC~613, NGC~1792, NGC~3621 and NGC~5734) the detection of the [\ion{Ne}{5}] lines is the only evidence of AGN activity. We decided to keep their \HII\ classification because: (1) the AGN may be extremely obscured or very low luminosity and, thus, the nuclear spectra might be dominated by star formation features; and (2) these lines are also detected in SNR, PN and WR stars which could be the origin of the [\ion{Ne}{5}] emission in these galaxies.

Table \ref{tbl_ratios} gives the median and deviation of all the line ratios we discuss in this paper for each type of galaxy. The line fluxes for all the galaxies are listed in Table \ref{tbl_fluxes_sample}.

\section{The Mid-IR High-Ionization Emission Lines}\label{s:o4ne5}

\subsection{The [\ion{O}{4}] versus [\ion{Ne}{5}] correlations}

The \Nevb\ and the \Oiv\ emission lines are commonly detected in active galaxies \citep{Lutz1998ULIRG,Genzel1998,Tommasin2010}. The detection of the former is considered a good indicator of AGN activity because of its high ionization potential. However using the \Oiv\ line as an AGN tracer is not as straightforward. For optically classified Seyfert galaxies the \Oiv\ luminosity is a good proxy for the AGN intrinsic luminosity \citep{Diamond2009, Melendez2008} and there is a good correlation between the \Nevb\ and the \Oiv\ emission in AGNs \citep{Dudik2009, Weaver2010}.
However, the \Oiv\ line is also produced by WR stars \citep{Schaerer1999} and is observed in starburst galaxies with no other evidence for the presence of an AGN \citep{Lutz1998,Bernard-Salas2009}.

\begin{turnpage}

\begin{figure*}
\includegraphics[width=0.43\textwidth]{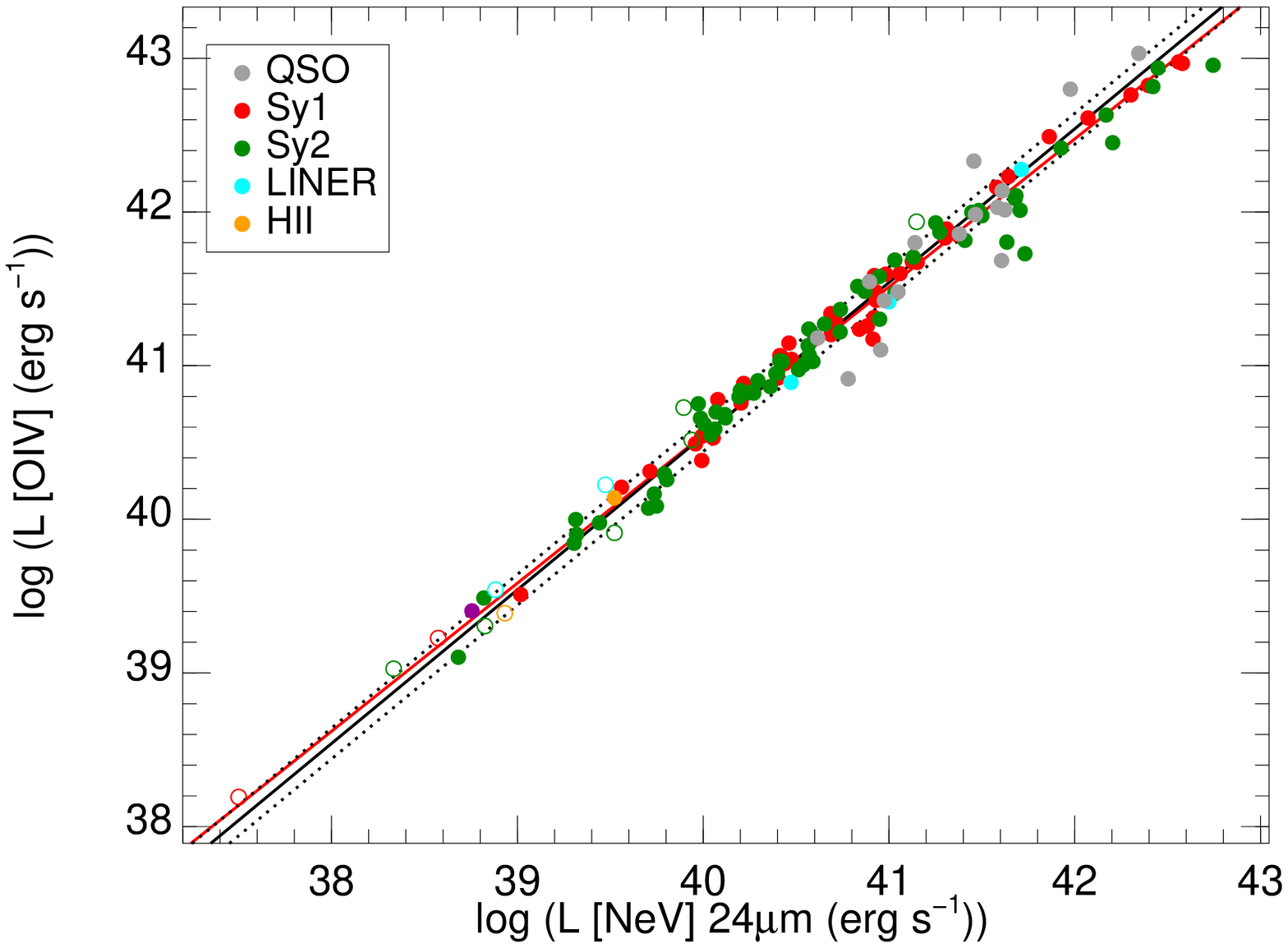}
\includegraphics[width=0.43\textwidth]{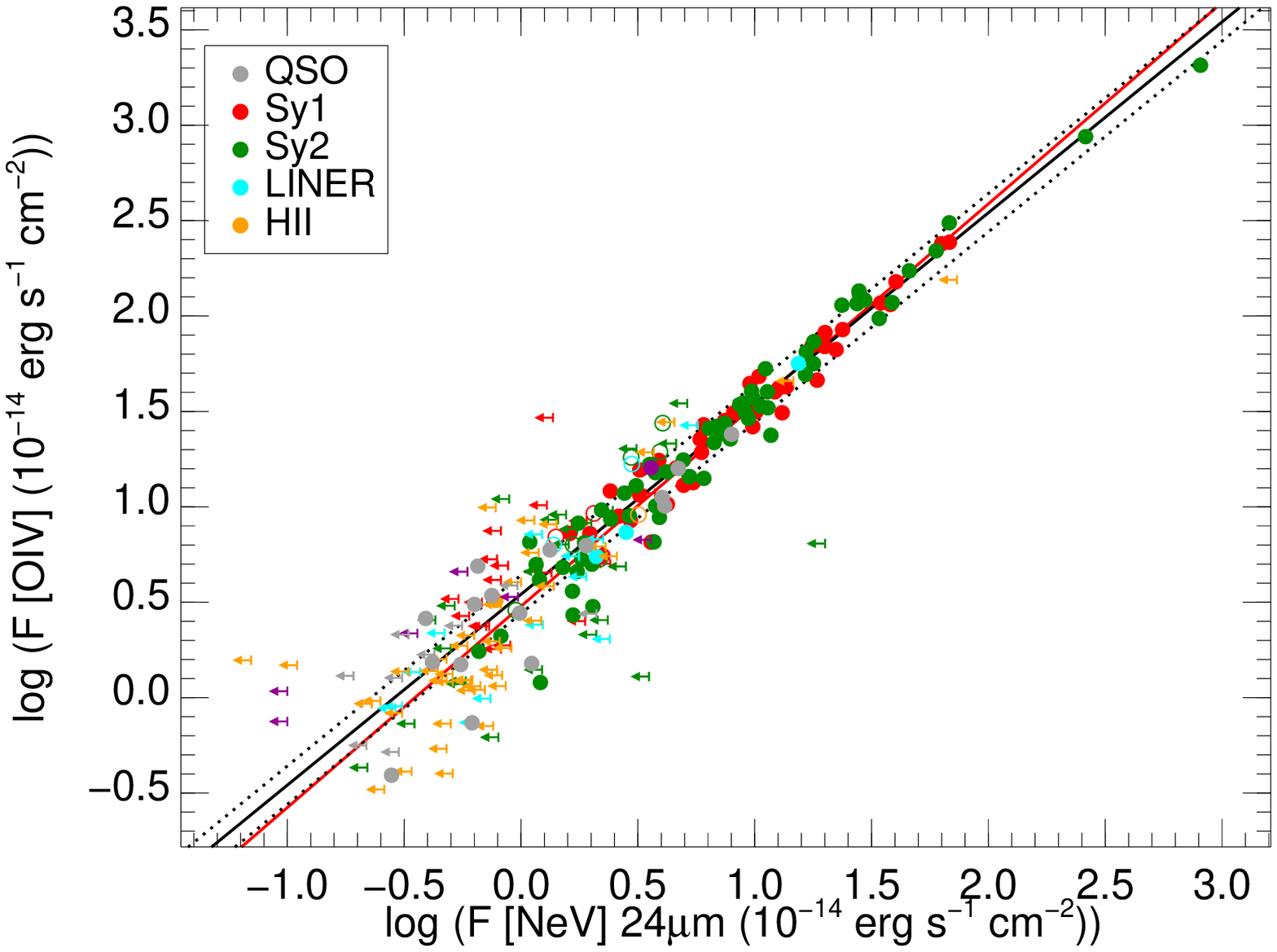}
\includegraphics[width=0.43\textwidth]{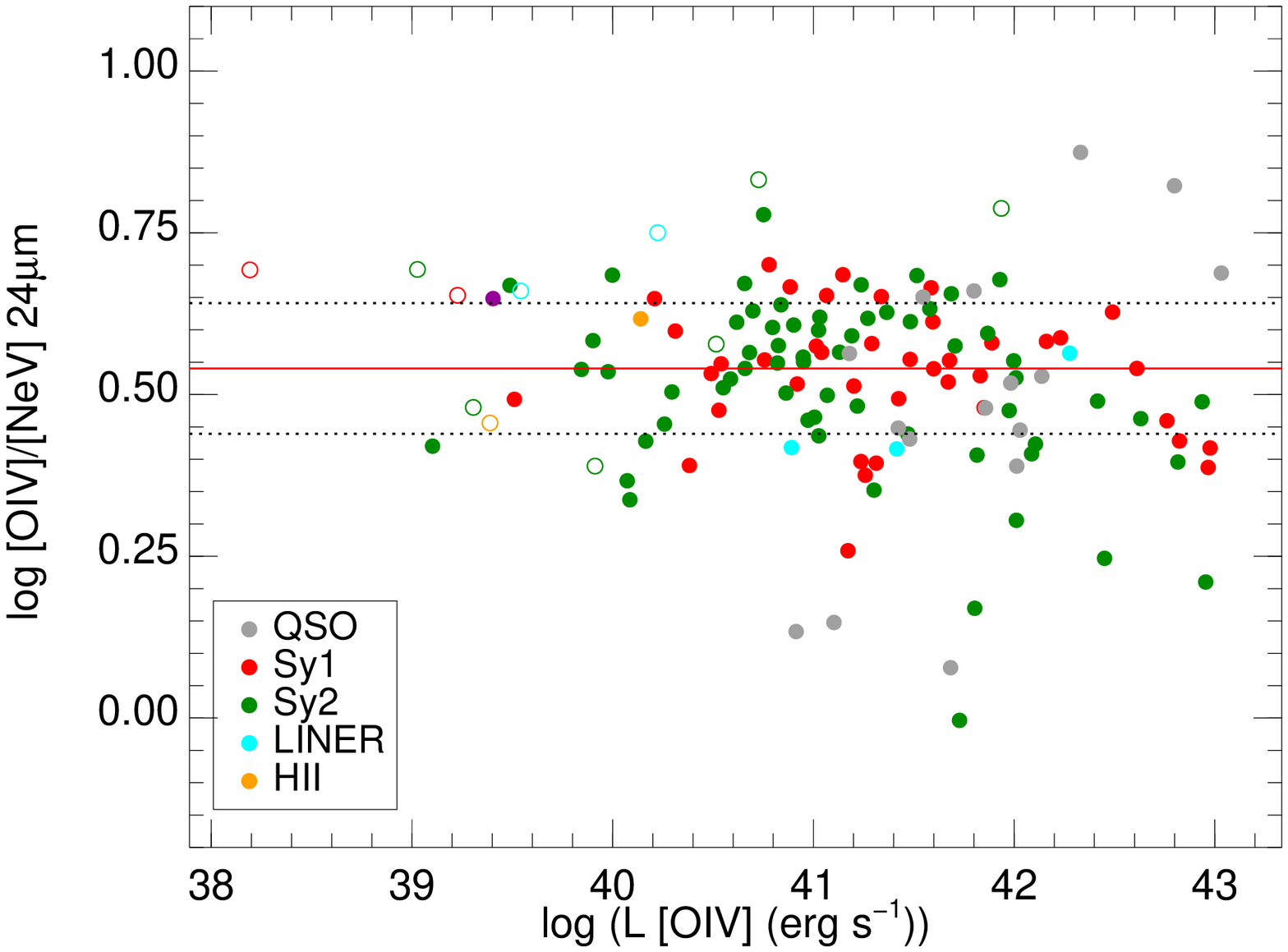}
\caption{Relationship between the \Oiv\ and \Nevb\ luminosities (left) and fluxes (middle).
The red line is the best fit and the black line the best linear fit. The dashed lines mark the 1$\sigma$ deviation. QSO are plotted as gray circles, Sy1 as red circles, Sy2 as green circles, LINERs as blue circles, \HII\ galaxies as orange circles and those galaxies without a classification as purple circles. The open symbols mark the galaxies with star formation which may contribute to the \Oiv\ emission (see Section \ref{s:ne3_agn} and right panel of Figure \ref{fig_ne5ne3}). The right panel shows the \Oiv\slash\Nevb\ ratio vs. the \Oiv\ luminosity. The solid red line is the ratio obtained from the linear fit and the dashed lines mark the 1$\sigma$ deviation.
}
\label{fig_ne5o4}
\vspace{10pt}
\includegraphics[width=0.43\textwidth]{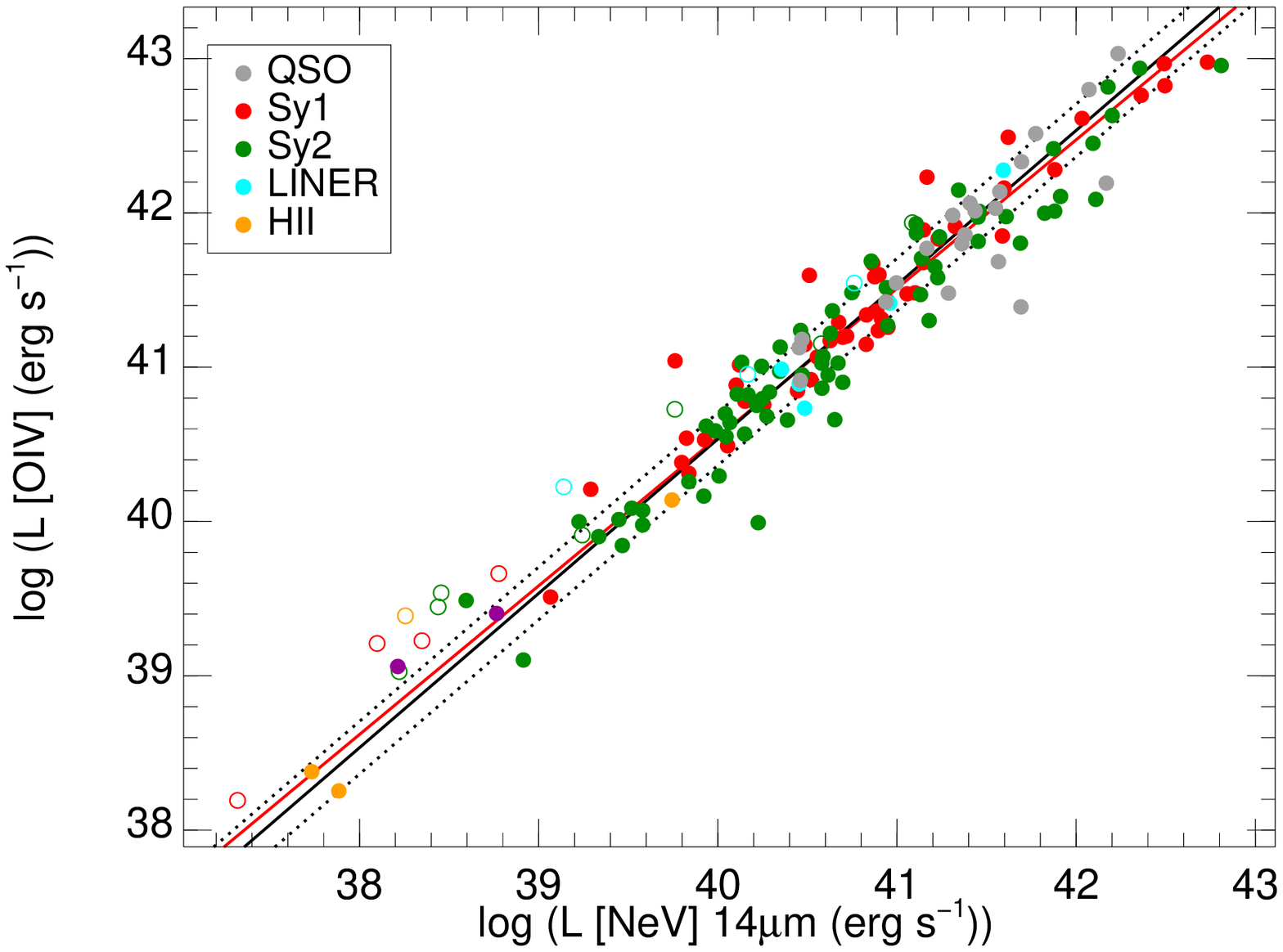}
\includegraphics[width=0.43\textwidth]{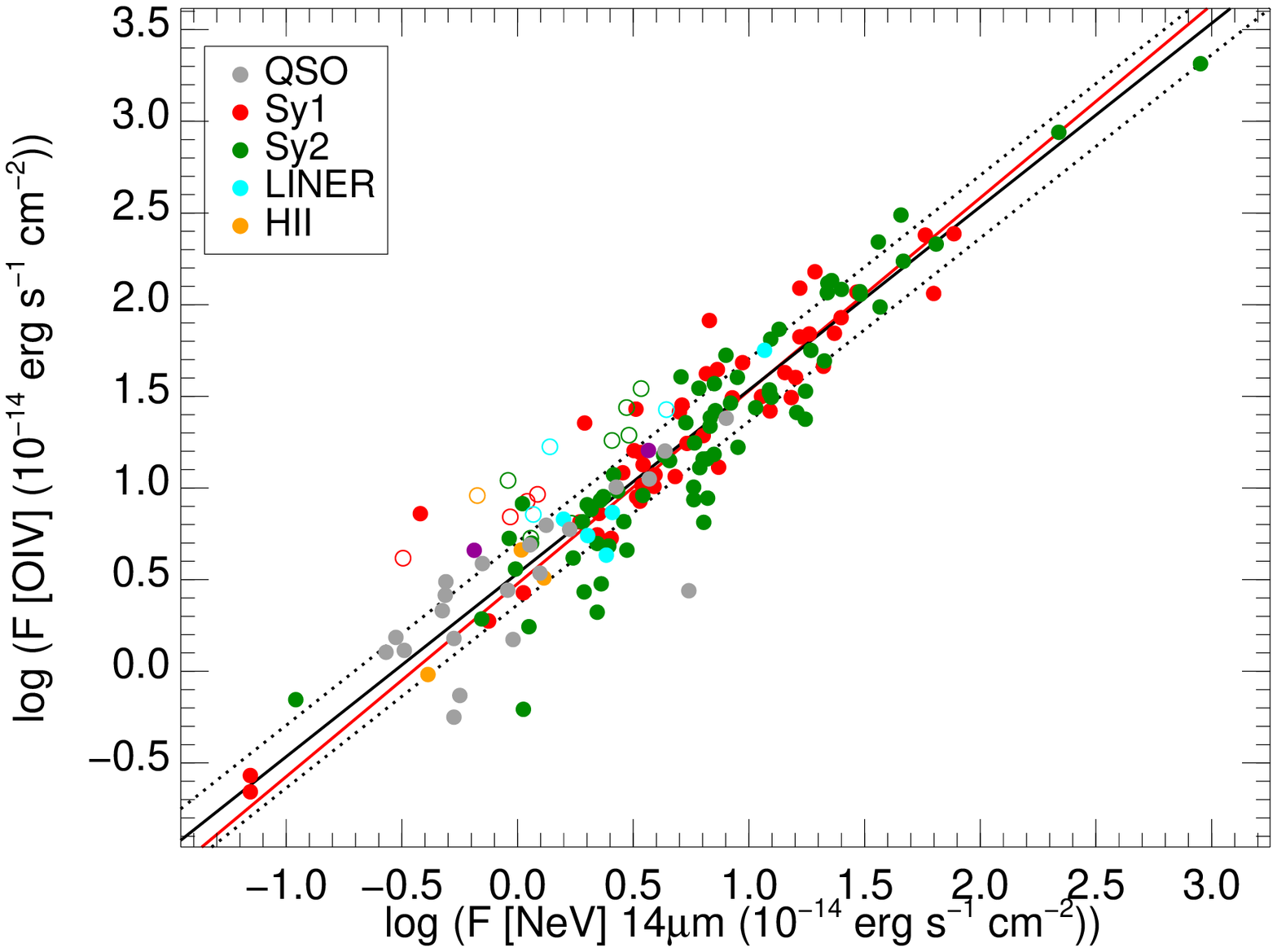}
\includegraphics[width=0.43\textwidth]{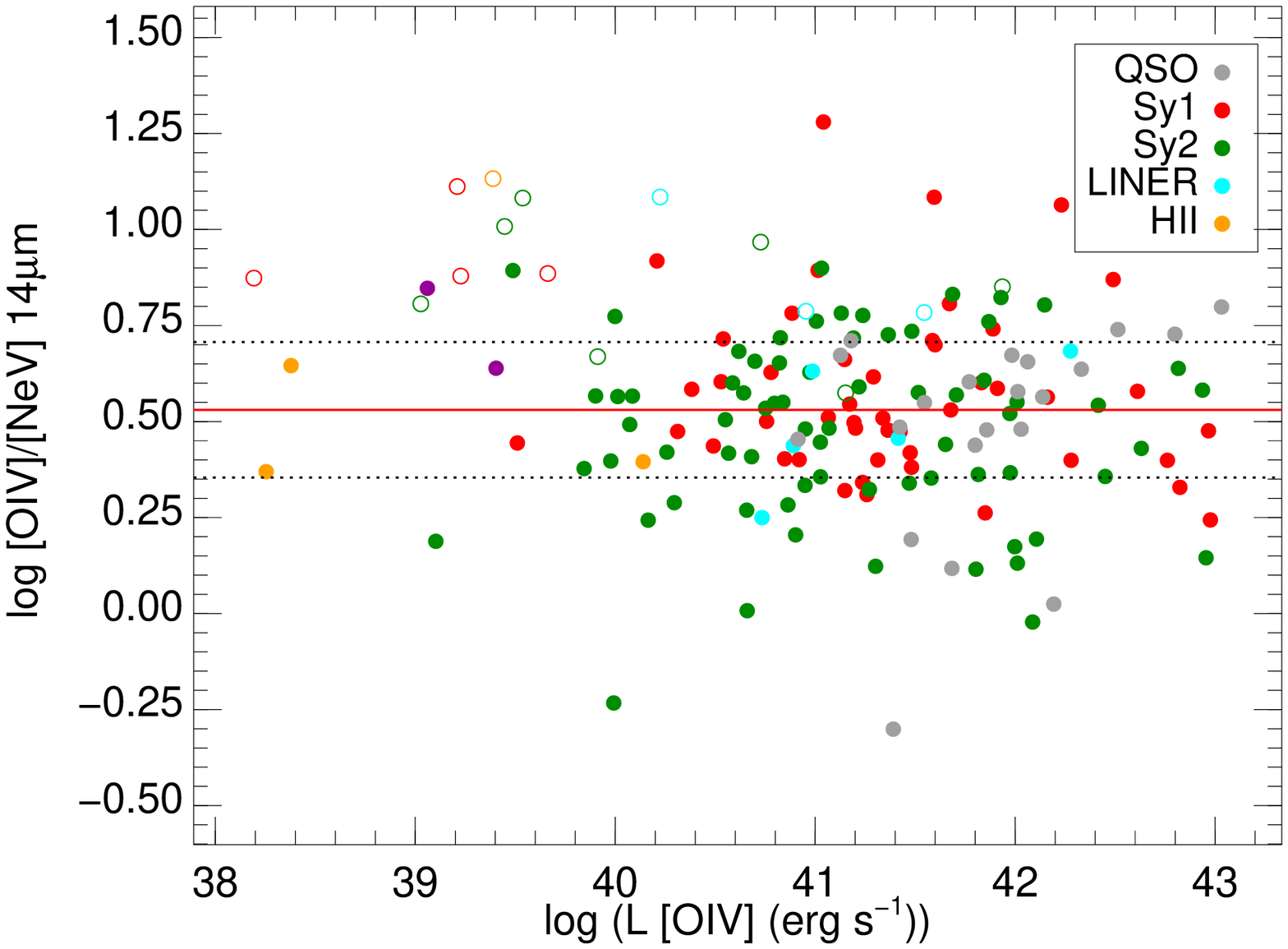}
\caption{Relationship between the \Oiv\ and \Neva\ luminosities (left) and fluxes (middle). The right panel shows the \Oiv\slash\Neva\ ratio vs. the \Oiv\ luminosity. Symbols are as in Figure \ref{fig_ne5o4}.}
\label{fig_ne5o4_2}
\end{figure*}

\end{turnpage}

The left panel of Figure \ref{fig_ne5o4} shows the tight correlation between the \Nevb\ and \Oiv\ lines spanning at least 5 orders of magnitude in luminosity (38 $<$ log $L_{\rm [O~{\scriptscriptstyle IV}]}$\,(erg s$^{-1}$) $<$ 43).
We do not find any dependence between the optical classification of the nuclear activity and this correlation. This implies that the ionization parameter is similar in type 1 and type 2 Seyfert galaxies (see Section \ref{s:models}). The slope of the best fit to the luminosity data is 0.96 $\pm$ 0.02 with a 0.14\,dex dispersion. It is reasonable to assume a linear correlation between these two emission lines, and this yields a ratio \Oiv\slash\Nevb\ $=$3.5, with rms scatter of 0.8.

The \Nevb\ line is detected in $\sim$50\% of the optically classified Seyfert galaxies (see Table \ref{tbl_sample}). However, for a considerable number of galaxies we have the \Oiv\ measurement and an upper limit for the \Nevb\ flux. We plot these fluxes and upper limits (middle panel of Figure \ref{fig_ne5o4}) for all the galaxies. The upper limits to the \Nevb\ flux are compatible with the correlation for most of the Seyfert galaxies. 
However, the \Oiv\ line is detected in more than 50\% of the \HII\ galaxies in our sample. In 90\% of these galaxies, the sensitivity of the \Nevb\ measurement (or upper limits) is inadequate to probe whether the [\ion{O}{4}] emission is associated with a hidden AGN.

In short, we find that the \Oiv\ luminosity is well correlated with the intrinsic AGN luminosity (i.e., [\ion{Ne}{5}] luminosity) for Seyfert galaxies and quasars with $L_{\rm [O~{\scriptscriptstyle IV}]} > 10^{39}$\,erg s$^{-1}$ to, at least, $L_{\rm [O~{\scriptscriptstyle IV}]} \sim 10^{43}$\,erg s$^{-1}$. For lower luminosities the fraction of the \Oiv\ emission produced by star formation may be considerable. Thus the \Oiv\ may not be an accurate tracer of the AGN luminosity for galaxies with $L_{\rm [O~{\scriptscriptstyle IV}]} < 10^{39}$\,erg s$^{-1}$\citep{Goulding2009}.

The linear correlation also holds for the \Neva\ and the \Oiv\ luminosities (Figure \ref{fig_ne5o4_2}), although the dispersion is larger (\Oiv\slash\Neva\ $=$3.4, , with rms scatter of 1.4 assuming a linear fit). %

\subsection{The [\ion{Ne}{3}] and [\ion{Ne}{2}] emission lines}\label{s:ne3_agn}

\begin{turnpage}

\begin{figure*}
\includegraphics[width=0.43\textwidth]{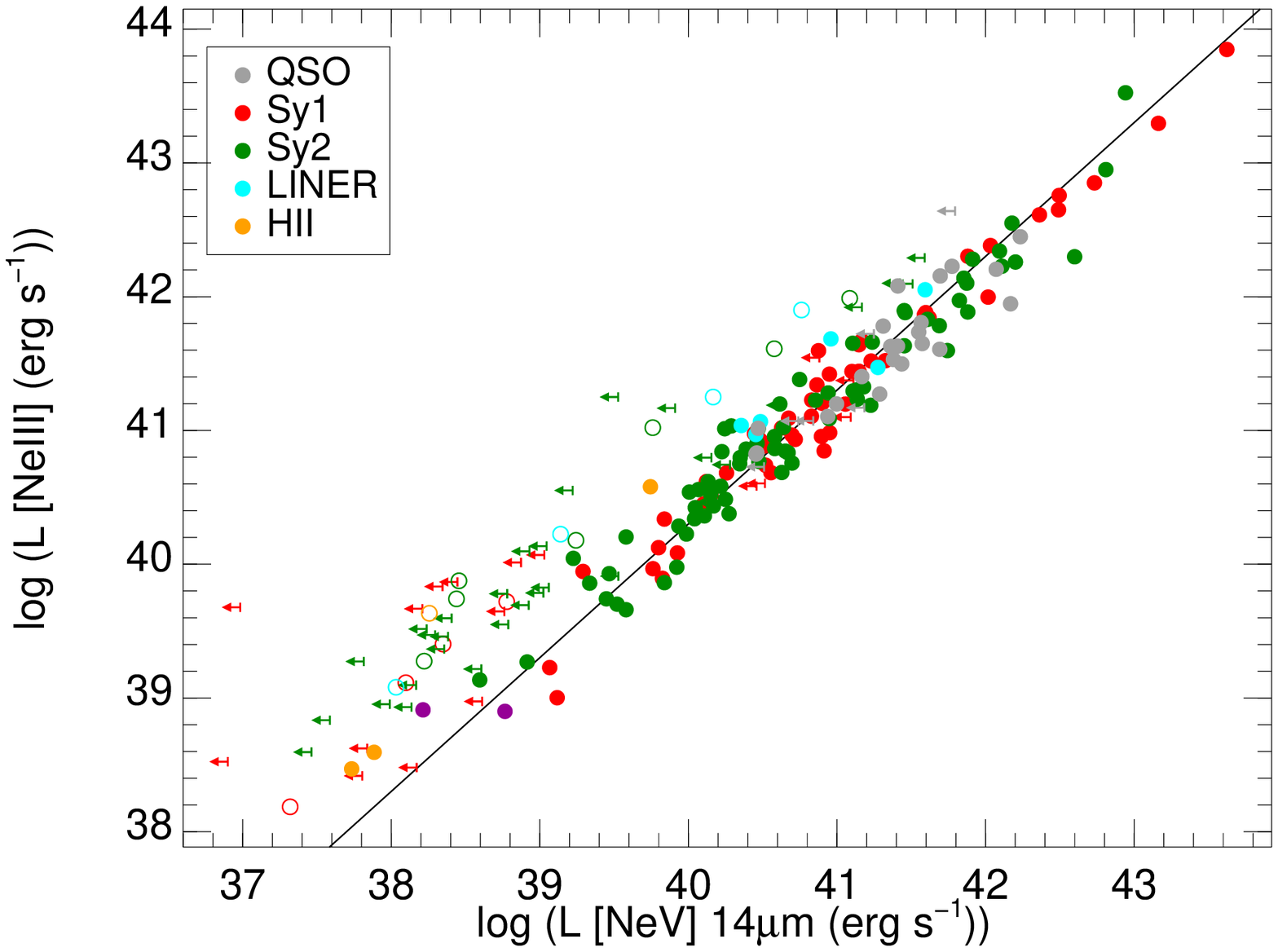}
\includegraphics[width=0.43\textwidth]{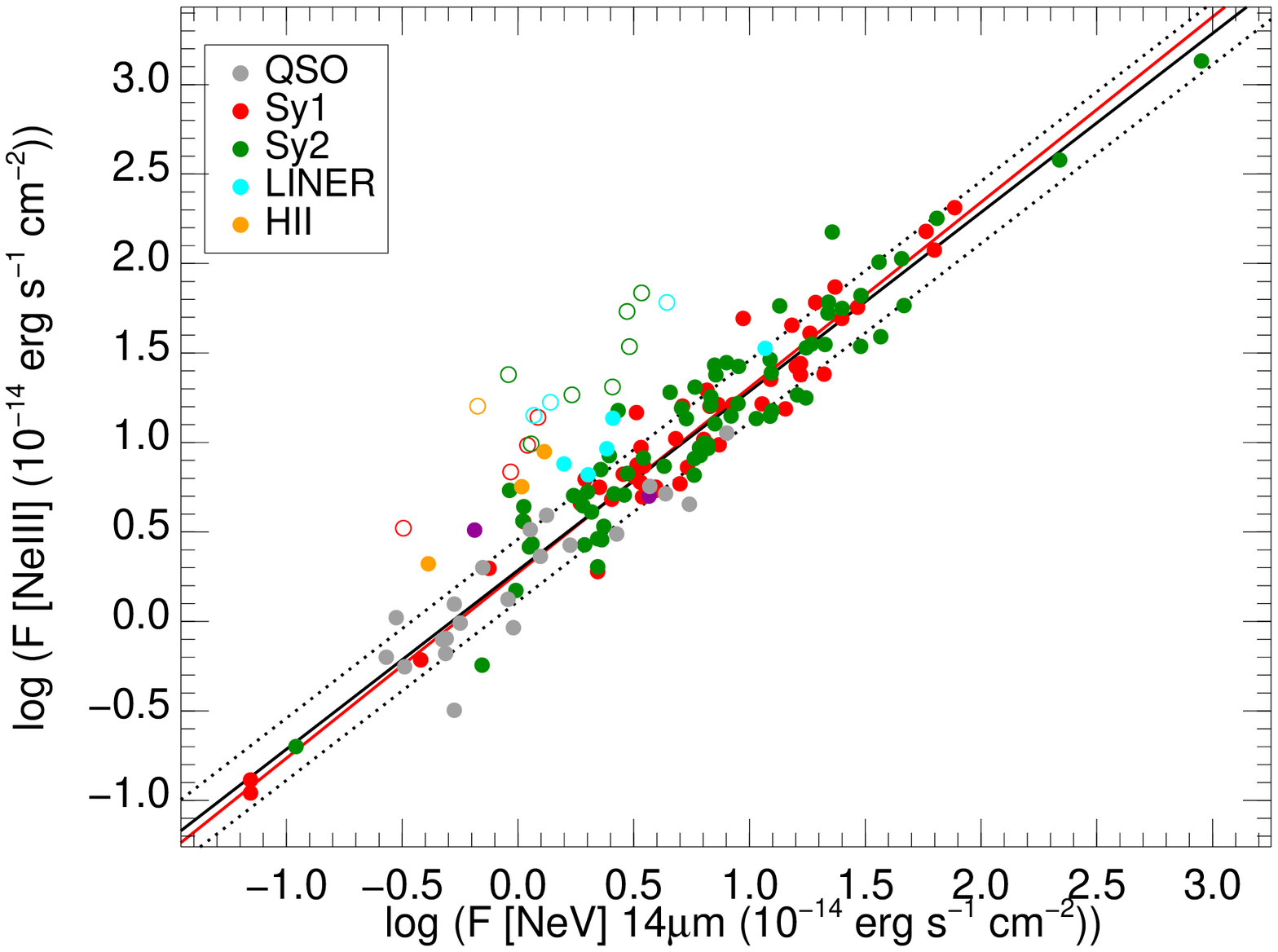}
\includegraphics[width=0.43\textwidth]{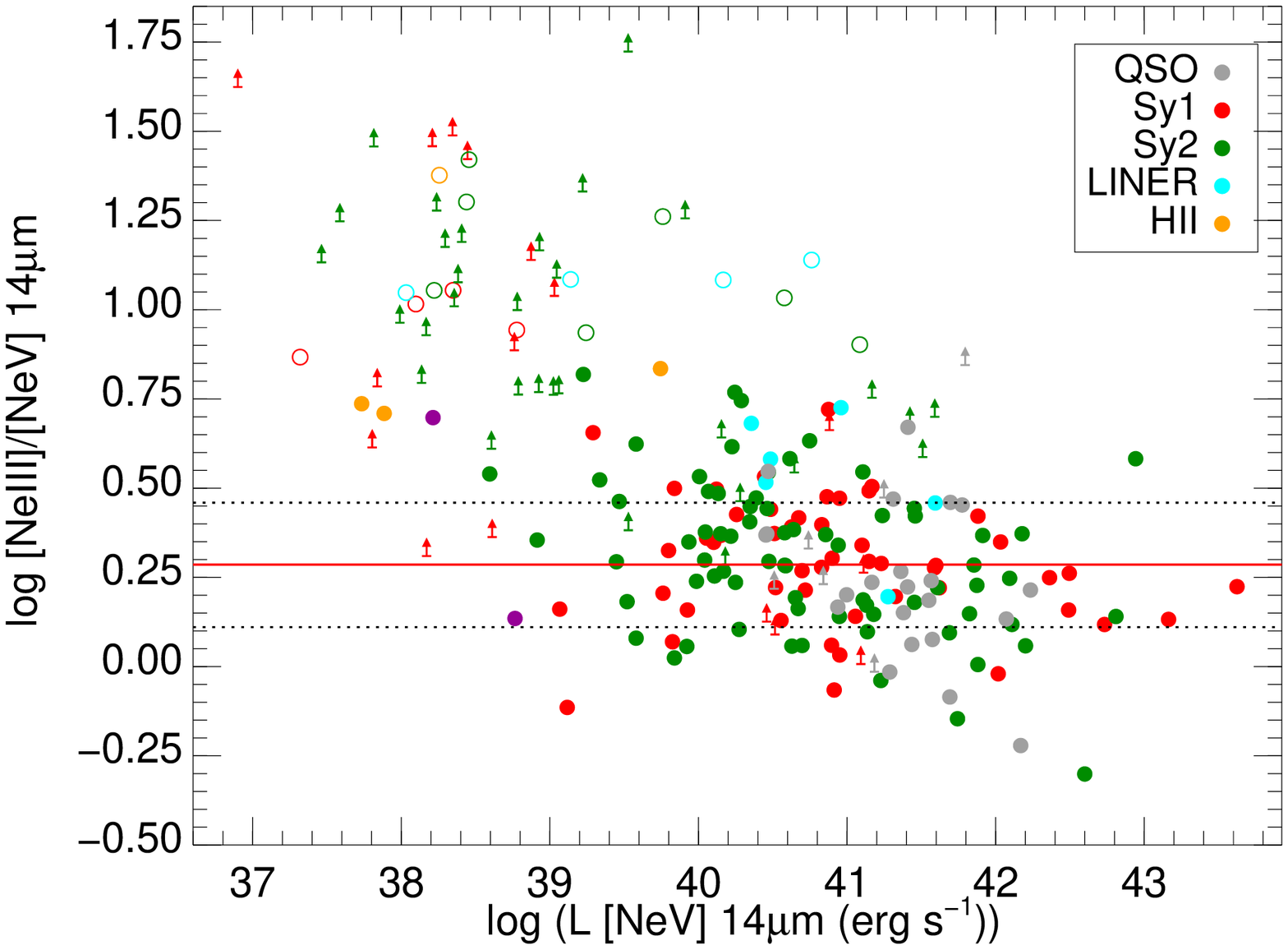}
\caption{Relation between the \Neiii\ and \Neva\ emission. Galaxy symbols are as in Figure \ref{fig_ne5o4}. The open symbols mark those galaxies above 3$\sigma$ the \Neva\ vs. \Neiii\ flux correlation. The black line in the left panel is the ratio obtained from the flux correlation. In the left panel only are included the \Nevb\ upper limits of the QSO and Seyfert galaxies.}
\label{fig_ne5ne3}
\end{figure*}

\end{turnpage}

\begin{figure}
\center
\includegraphics[width=0.45\textwidth]{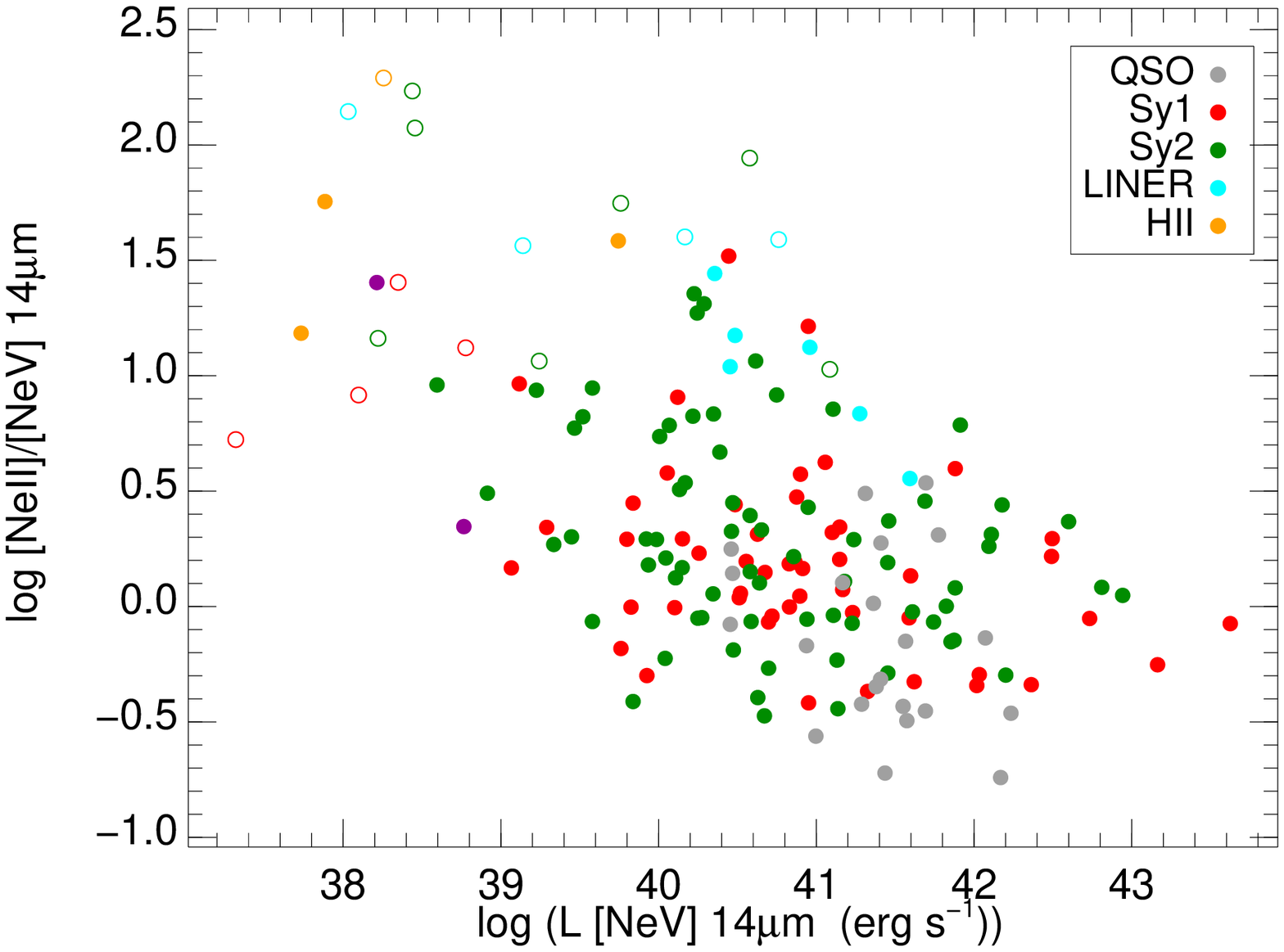}
\caption{\Neii\slash\Neva\ ratio vs. \Neva\ luminosity. Symbols are as in Figure \ref{fig_ne5ne3}.}
\label{fig_ne5ne2}
\end{figure}

Due to the intermediate ionization potential of Ne$^{+2}$ (41.0\,eV), the \Neiii\ emission line may be produced by young stars or by AGNs. This line is observed in star forming galaxies \citep{Verma03, Brandl06, Beirao2006, Beirao08, Ho07, Bernard-Salas2009, Dale2009, Pereira2010}, and it is also correlated with the AGN intrinsic luminosity \citep{Gorjian2007, Deo2007, Tommasin08, Melendez2008b}.

The middle panel of Figure \ref{fig_ne5ne3} shows the relation between the observed flux of the \Neiii\ and \Neva\ emission lines for our sample of AGNs and for the 4 \HII\ galaxies with [\ion{Ne}{5}] detections. We compare these lines because they are close in wavelength and the effect of differential extinction is minimized. It is clear that some galaxies have an excess of \Neiii\ emission relative to that of the \Neva, which we attribute to star formation (see Section \ref{s:ne3sf}). Therefore to obtain the fit to the data we use the outliers resistant linear fit algorithm provided by the IDL function ROBUST\_LINEFIT. The slope of this fit is 1.02\,$\pm$\,0.02  with a scatter of about 0.2 dex around the fit. The result is consistent with a simple correlation; by fixing the slope to unity, we obtain the ratio \Neiii\slash\Neva\ $=$1.9, , with rms scatter of 0.8. %
This correlation was previously reported by \citet{Gorjian2007}, although they found a slight dependence of the ratio with the \Neva\ luminosity (slope 0.89). Their sample included 53 X-ray selected AGNs with \Neva\ luminosities between 10$^{39}$ and 10$^{43}$\,erg s$^{-1}$. The larger number of galaxies in our sample allows us to minimize the contribution of the galaxies contaminated by star formation to the fit and the luminosity dependence of the correlation is substantially reduced. Where the AGN clearly dominates, the points fall around the line determined for the whole sample by ROBUST\_LINEFIT, thus supporting our interpretation that this fit determines the relationship for AGNs in general.

The \Neii\ emission traces young ($<$10\,Myr) star formation \citep{Roche1991, Thornley2000, Verma03, Rigby2004, Snijders07, Ho07, DiazSantos2010}. The ionization potential of this emission line is 21\,eV and thus it is mainly produced in \HII\ regions.
Figure \ref{fig_ne5ne2} shows that the scatter in the \Neii\slash\Neva\ ratio (0.7\,dex) is larger than that found in the other line ratios, for the lower AGN luminosities there is a large scatter. There is a weak trend of \Neii\slash\Neva\ ratios decreasing with increasing AGN luminosity. Assuming that there is an intrinsic ratio for AGN, this behavior implies that the AGN does not dominate the \Neii\ output and that there is a wide range of star formation rates contributing to its luminosity. However, for the most luminous objects ($\log L_{\rm [Ne~{\scriptscriptstyle V}]14\mu m}~({\rm erg~s^{-1}}) >$ 41.5), the smaller scatter indicates that the relative contribution of the AGN to the total \Neii\ excitation is higher.

\citet{Sturm2002} determined that the pure AGN \Neva\slash\Neii\ ratio is 1.1, but we find many Seyfert galaxies with higher ratios. The largest \Neva\slash\Neii\ ratio in our sample is $\sim$5.5, which is compatible with the largest ratio predicted by NLR photoionization models (\citealt{Groves2006}, their Figure 11).
With our data (Figure \ref{fig_ne5ne2}) we are not able to determine the pure AGN ratio, which also depends on the ionization parameter \citep{Groves2006}, but it is likely that it is between that found by \citet{Sturm2002} and the largest ratio in our sample.

\begin{figure}
\center
\includegraphics[width=0.45\textwidth]{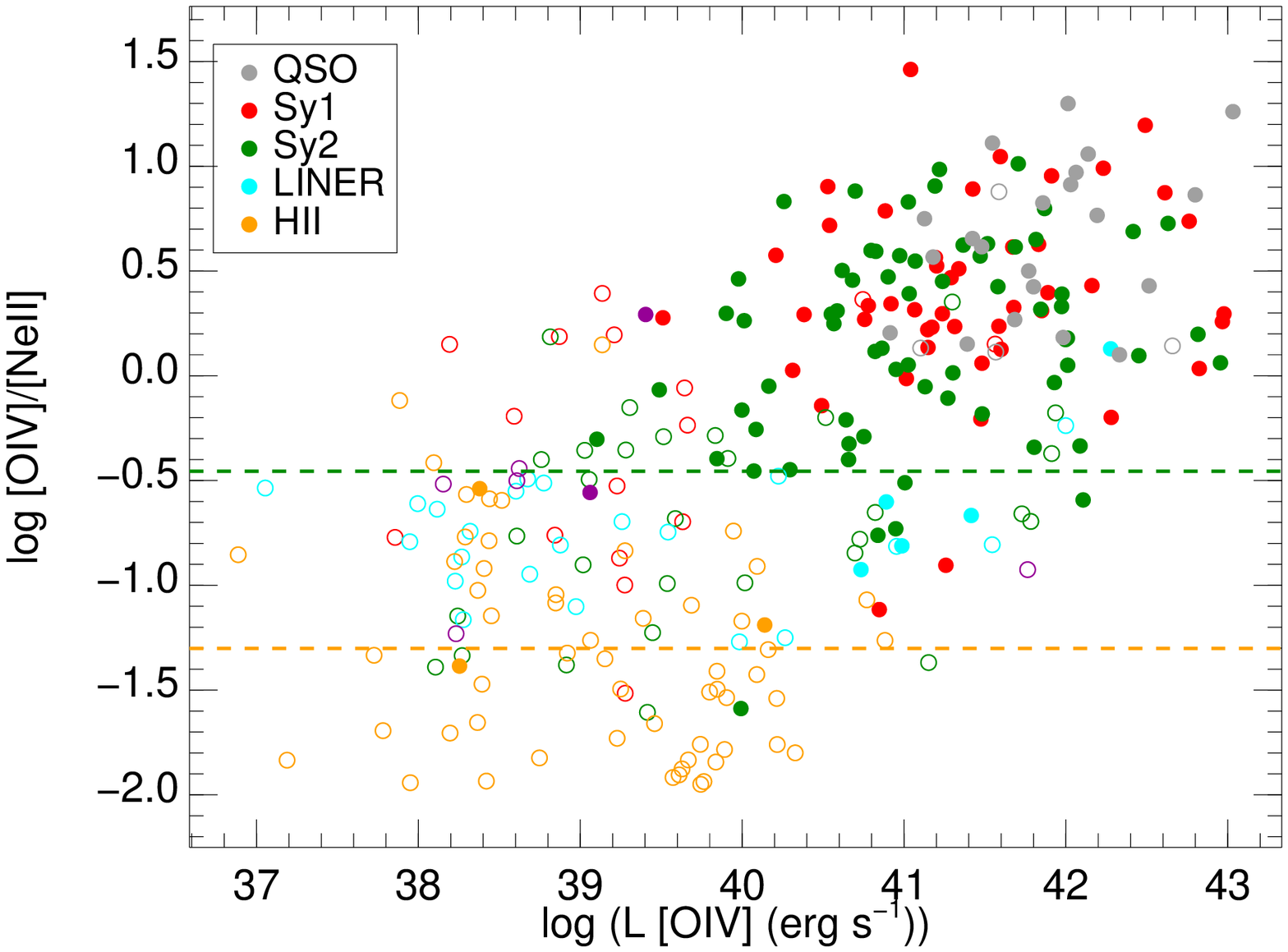}
\caption{\Oiv\slash\Neii\ ratio vs. \Oiv\ luminosity. Galaxy symbols are as in Figure \ref{fig_ne5o4}.
The dashed orange ([\ion{O}{4}]\slash[\ion{Ne}{2}] $=0.05$) and green ([\ion{O}{4}]\slash[\ion{Ne}{2}] $=0.35$) lines are the boundaries of the \HII\ and  Seyfert galaxies respectively.}
\label{fig_o4ne2}
\end{figure}

The \Oiv\slash\Neii\ ratio can be used to separate AGNs from star-forming galaxies \citep{Genzel1998,Sturm2002,Peeters2004,Dale2009}. We plot this ratio versus the \Oiv\ luminosity in Figure \ref{fig_o4ne2}. Most of the active galaxies have ratios larger than 0.35 whereas the ratios for \HII\ galaxies are lower than 0.05 (Figure 5 of \citealt{Dale2009}). LINERs appear in the region between these ratios (Figure \ref{fig_o4ne2}). A considerable number of \HII\ galaxies also have \Oiv\slash\Neii\ ratios above 0.05. The [\ion{Ne}{5}] lines are not detected for them although their upper limits are compatible with the \Oiv\ versus [\ion{Ne}{5}] correlations and thus a low-luminosity AGN could be the origin of the \Oiv\ emission.
The large scatter in the \Oiv\slash\Neii\ ratio, due to the different star formation rates in the AGNs, does not allow us to determine the value of the pure AGN ratio. The largest \Oiv\slash\Neii\ ratios that we find in AGNs are $\sim$10.

\section{Star formation contributions to the Mid-IR High-Ionization Emission Lines}\label{s:ne3sf}

\subsection{[\ion{O}{4}] Contamination by Star Formation}\label{ss:o4contamination}

In Figures \ref{fig_ne5o4} and \ref{fig_ne5o4_2} there are few outliers above the \Oiv\ versus [\ion{Ne}{5}] correlations.
Most of them are represented by open symbols, which indicate that their \Neiii\slash\Neva\ ratios (see Section \ref{ss:ne3sf_estimation}) are larger than those found in AGN dominated galaxies (QSO, Sy1 and Sy2). It is likely that the \Oiv\ emission of these galaxies is contaminated by star formation.
We now compare the \Oiv\ emission of star forming regions and AGN. The \Oiv\slash\Neii\ ratio in star forming regions is $<$ 0.05 (Section \ref{s:ne3_agn}). Taking this upper limit and using the relation between the \Neii\ luminosity and the total infrared (8-1000\micron) luminosity ($L_{\rm IR}$) from \citet{Ho07} we obtain the following relation for star-forming galaxies.

\begin{equation}\label{eqn:o4lir}
\log L_{\rm [O~{\scriptscriptstyle IV}]}~({\rm erg~s^{-1}})< \log L_{\rm IR}~({\rm erg~s^{-1}}) - 4.7 \pm 0.6
\end{equation}

\citet{Rigby2009} calculated the ratio between the total AGN luminosity and the \Oiv\ emission. We use the ratio for type 1 AGNs ($\sim$2500) because the hard X-ray (E $>$ 10\,keV) emission used for the bolometric corrections may be affected by extinction in Seyfert 2s \citep{Rigby2009}.

\begin{equation}\label{eqn:lo4lagn}
\log L_{\rm [O~{\scriptscriptstyle IV}]}~({\rm erg~s^{-1}}) = \log L_{\rm AGN}~({\rm erg~s^{-1}}) - 3.4 \pm 0.4
\end{equation}

Then, combining both relations we find that the luminosity due to star formation, $L_{\rm IR}$, has to be at least 20 times brighter than the AGN luminosity (that is, AGN contribution to the total luminosity below 5\%) for the star formation to dominate the \Oiv\ emission.

We note that the number of \HII\ galaxies falls rapidly above log $L_{\rm [O~{\scriptscriptstyle IV}]}$\,(erg s$^{-1}$) $=$ 40.2 (see Figure \ref{fig_o4ne2}). Assuming that star formation dominates the \Oiv\ emission at this luminosity in these galaxies, the equivalent infrared luminosity ($L_{\rm IR}$) would be 2$\times$10$^{11}$\,$L _{\rm \sun}$. This value of the $L_{\rm IR}$ seems reasonable since most of the \HII\ galaxies with log $L_{\rm [O~{\scriptscriptstyle IV}]}$\,(erg s$^{-1}$) between 39.6 and 41.0 are classified as LIRGs.
The $L_{\rm IR}$ limit calculated above is one order of magnitude larger than $L_{\rm \star}$ \citep{Takeuchi2003} so the reason for the lower number of \HII\ galaxies above this \Oiv\ luminosity can be that the space density of luminous star forming galaxies decreases rapidly with increasing luminosity.

\begin{figure}[h]
\center
\includegraphics[width=0.45\textwidth]{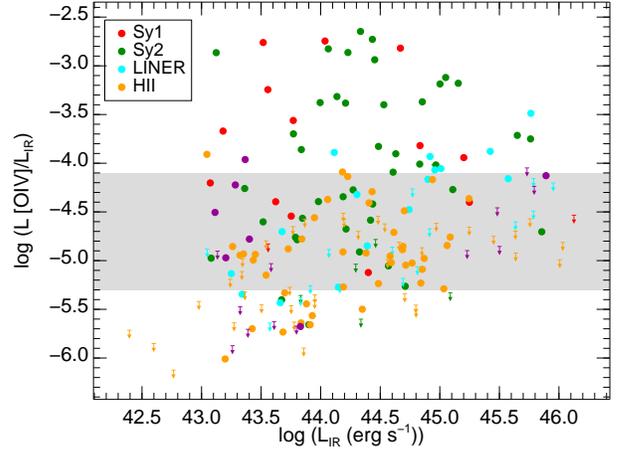}
\caption{\Oiv\slash $L_{\rm IR}$ ratio vs. $L_{\rm IR}$ luminosity. Galaxy symbols are as in Figure \ref{fig_ne5o4}. The shaded region is the upper limit to the \Oiv\slash $L_{\rm IR}$ ratio of star-forming galaxies. Galaxies above this region need an AGN contribution to the \Oiv\ emission in order to explain the observed ratio.
}
\label{fig_lo4_lir}
\end{figure}

Out of 196 galaxies with estimates for $L_{\rm IR}$, \Oiv\ is detected in 122.
Only 2\,$\pm$\,2\,\% of the \HII\ galaxies have \Oiv\ luminosities \hbox{1$\sigma$} above those expected from their $L_{\rm IR}$. Likewise star-formation can explain the observed \Oiv\ luminosities for 9\,$\pm$\,4\,\% of the Seyfert galaxies (Figure \ref{fig_lo4_lir}).
Thus the \Oiv\ emission of Seyferts is generally dominated by the AGN, whereas for optically classified \HII\ galaxies star formation is likely to be the origin of the \Oiv\ emission.

\subsection{The [\ion{Ne}{3}] Excess}\label{ss:ne3sf_estimation}

\begin{figure}
\center
\includegraphics[width=0.45\textwidth]{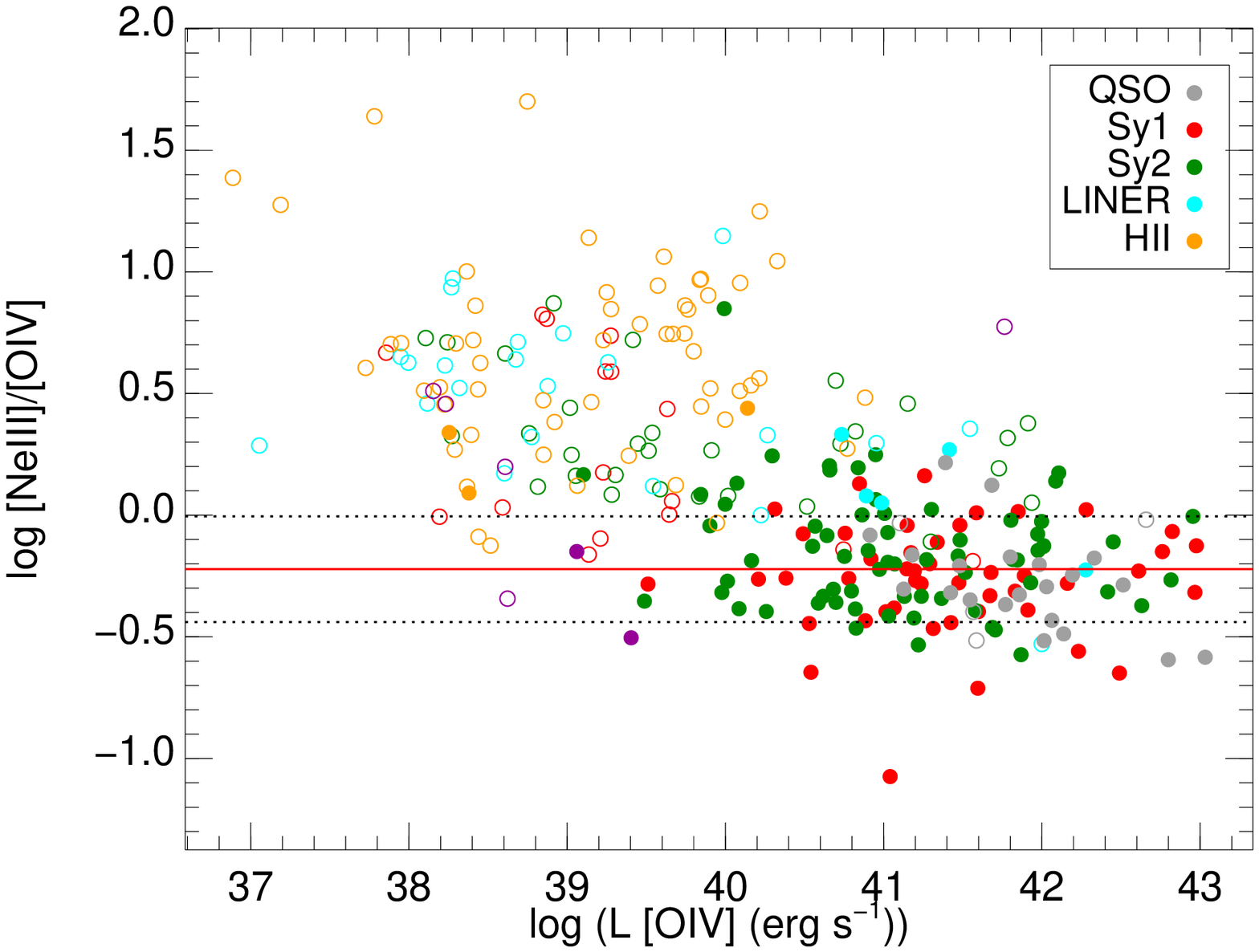}
\caption{\Neiii\slash\Oiv\ ratio vs. \Oiv\ luminosity. Galaxy symbols are as in Figure \ref{fig_ne5o4}. The solid red line is the \Oiv\slash\Neiii\ ratio calculated from the linear fit to the \Oiv\ vs. \Neiii\ luminosities. The dotted black lines are the 1$\sigma$ deviation.
}
\label{fig_lo4_o4ne3}
\end{figure}

The \Neiii\ emission may be produced by star formation and/or AGN. In this section we try to quantify the contribution of each one to the total \Neiii\ output.
For this purpose we used the correlations between \Neva, \Oiv\ and \Neiii\ (see Section \ref{s:o4ne5}). Combining the typical \Oiv\slash\Neva\ and \Neiii\slash\Neva\ ratios found in AGN we estimate a ratio \Neiii\slash\Oiv\ $=$ 0.6, with rms scatter of 0.3 for AGNs\footnote{In this section the term ``AGN'' refers to galaxies classified as QSO or Seyfert.}, or log [\ion{Ne}{3}]\slash[\ion{O}{4}] $=$ \hbox{--0.22}. This estimated ratio is very similar to that obtained using a linear fit between the \Neiii\ and \Oiv\ luminosities (Table \ref{tbl_ratios}).
The median of this ratio for \HII\ galaxies is 4.7, or log [\ion{Ne}{3}]\slash[\ion{O}{4}] $=$ 0.67, which is $\sim$8 times larger than that estimated for active galaxies.
Figure \ref{fig_lo4_o4ne3} shows that the observed ratio for Seyfert galaxies and QSOs falls around the predicted value for AGN. Only a small fraction (4\%) of the \HII\ galaxies have \Neiii\slash\Oiv\ ratios in the AGN range.
It is also apparent that \HII\ galaxies, a large fraction of LINERs and some Seyfert galaxies have an excess of \Neiii\ 
relative to their \Oiv\ emission which can also be attributed to star formation.

We find that the four \HII\ galaxies with \Neva\ detections have the \Neiii\slash\Neva\ and \Neii\slash\Neva\ ratios much larger than those of Seyfert galaxies (Table \ref{tbl_ratios}). This indicates that the star-formation contribution is larger than that in Seyfert galaxies. Thus as suggested in Section \ref{s:sample} the AGN does not dominate the nuclear spectra of these galaxies.

As can be seen in the left panel of Figure \ref{fig_ne5ne3}, most of the outliers (galaxies more than 3$\sigma$ above the \Neva\ versus \Neiii\ flux correlation) have $L_{\rm [Ne~{\scriptscriptstyle III}]} < 10^{42}$\,erg s$^{-1}$. That is, for the most luminous objects the AGN dominates the \Neiii\ emission. Using the relation by \citet{Ho07} that relates star formation rate (SFR) and the luminosity of the \Neii\ and \Neiii\ emission lines and assuming a \Neiii\slash\Neii\ ratio $=0.3$ (see Section \ref{ss:ne3sf_ratio}) we find that the SFR of the outliers is between 0.1 and 100\,$M_\sun$ yr$^{-1}$.

\subsection{The \Neiii\slash\Neii\ Ratio}\label{ss:ne3sf_ratio}

\begin{figure}
\center
\includegraphics[width=0.45\textwidth]{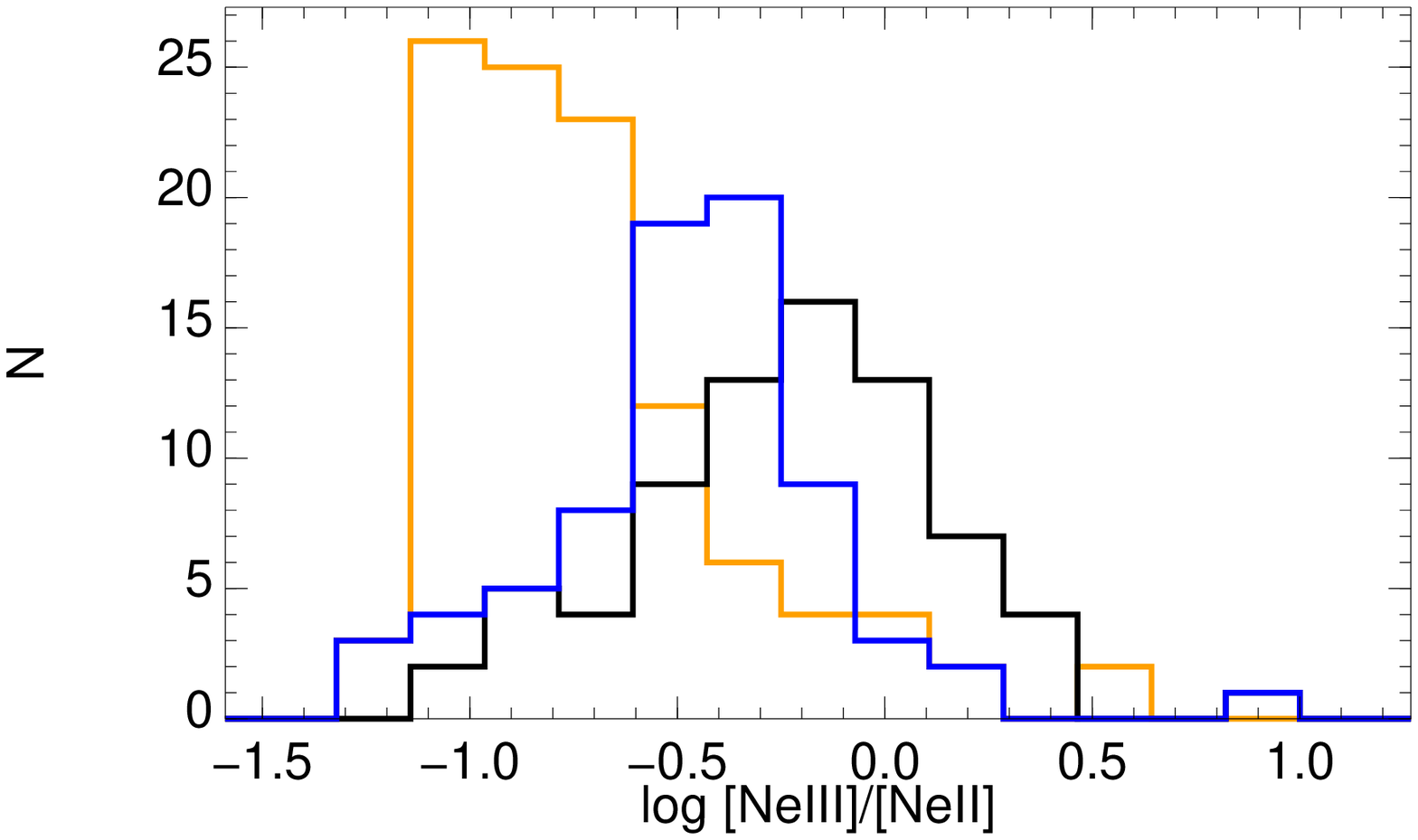}
\caption{Distribution of the \Neiii\slash\Neii\ ratio. The orange histogram corresponds to the observed ratios in \HII\ galaxies, black corresponds to the observed ratios in active galaxies with star formation (\Neiii\slash\Oiv\ ratios $>$ 0.9). The blue histogram shows the ratios calculated for Seyfert galaxies after subtracting the AGN contribution to the \Neiii\ emission.}
\label{fig_ne3ne2_histo}
\end{figure}

The \Neiii\slash\Neii\ ratio traces the hardness of the radiation field and the age of the stellar population \citep{Rigby2004, Snijders07}.
Figure \ref{fig_ne3ne2_histo} compares the \Neiii\slash\Neii\ ratio observed in \HII\ galaxies (median 0.2) with that observed in Seyfert galaxies with star formation\footnote{those with excess \Neiii.} (median 0.6). The larger ratio found in Seyfert galaxies can be explained if the AGN contributes noticeably to the total \Neiii\ emission but not to the \Neii\ emission (see Section \ref{s:ne3_agn}).
By assuming this we calculate the star formation \Neiii\slash\Neii\ ratio in these Seyfert galaxies using the non-AGN \Neiii\ emission that we estimated using the method described in Section \ref{s:ne3sf_sy1sy2}. The median of this ratio becomes $\sim$0.3, which is somewhat larger than that found in pure \HII\ galaxies but it is in the range of the observed ratio in this class of galaxies \citep{Brandl06}.

\section{Comparison of Star Formation in Seyfert 1 and 2 Galaxies}\label{s:ne3sf_sy1sy2}

\begin{figure*}
\center
\includegraphics[width=.95\textwidth]{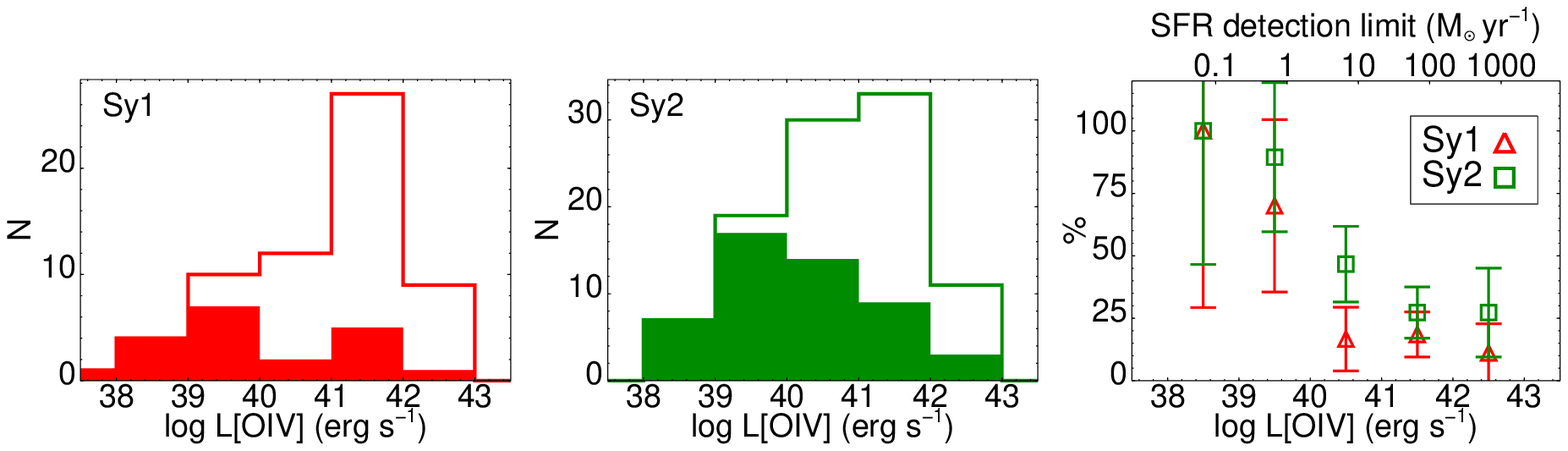}
\caption{Distribution of the \Oiv\ luminosities of the Seyfert 1 (left) and 2 (middle) galaxies. The filled histograms are the galaxies with a \Neiii\ excess. The right panel compares the fraction of type 1 and type 2 Seyferts with a \Neiii\ excess for each luminosity bin. We used Poisson statistical errors to estimate the uncertainties of the fractions. The scale on the upper part of the right panel gives the lower limit to the SFR that can be detected with our method for each AGN luminosity bin.}
\label{fig_comparison_sy1sy2_all}
\end{figure*}

The AGN unification scenario predicts that there should be no differences between the star formation activity in Seyfert 1 and 2 galaxies.
A number of studies compared the star formation rates in Seyfert 1 and 2 galaxies. Some of them found enhanced star formation activity in Seyfert 2 with respect Seyfert 1 galaxies, for a given AGN luminosity \citep{Maiolino1995AGN, Buchanan2006, Deo2007, Melendez2008b}, whereas
others did not find any differences between the two types \citep{Kauffmann2003, Imanishi2004}.
However, as discussed by \citet{Shi2009}, the different methods used to study the star formation activity are sensitive to different stellar age ranges. Therefore these apparently contradictory results may be consistent with each other.
The \Neiii\ line is sensitive to young ($<$5\,Myr) massive star formation \citep{Rigby2004,Snijders07}. Studies using this line (and the \Neii\ line) found that star formation is enhanced in Seyfert 2 galaxies \citep{Deo2007, Melendez2008b}.
\citet{Kauffmann2003} did not find differences between Seyfert 1 and Seyfert 2 star formation, but their method (the 4000\AA\ break and the H$\delta$ stellar absorption) was sensitive to stellar populations older than 100 Myr \citep{Shi2009}.
The same applies to the PAH features used by \citet{Imanishi2004}, which can be affected by the presence of the AGN \citep[][and references therein]{Diamond2010}.

It is also important to use the appropriate indicator for the AGN luminosity. 
We will use the \Oiv\ emission, as it is a good isotropic indicator \citep{Rigby2009, Diamond2009}, as long as the AGN dominates the \Oiv\ emission as we showed in Section \ref{ss:o4contamination}.

We estimate the star formation contribution to the \Neiii\ emission by using the observed \Oiv\slash\Neiii\ ratio and subtracting the estimated AGN contribution from the total \Neiii\ line strength. Further details are provided in Appendix \ref{apxContribution}.
We used the \Oiv\slash\Neiii\ instead of the \Oiv\slash\Neii\ ratio because we could not calculate the typical AGN \Oiv\slash\Neii\ ratio due to the larger relative contribution of star formation activity to the \Neii\ emission (Section \ref{s:ne3_agn}).
We only subtract this contribution from those active galaxies more than 1$\sigma$ above the typical AGN \Neiii\slash\Oiv\ ratio (see Section \ref{ss:ne3sf_estimation} and Figure \ref{fig_lo4_o4ne3}), since the galaxies below this ratio are presumably dominated by the AGN and their star formation activity, if any, is masked. This criterion basically selects only those active galaxies with at least 25\% of the \Neiii\ emission arising from star formation. In our sample 45\% of the AGNs with the \Oiv\ line detected have excess \Neiii\ emission due to star formation.
The right panel of Figure \ref{fig_comparison_sy1sy2_all} shows that the fraction of Seyfert 2s with star formation is slightly higher than that of Seyfert 1s.
In all the luminosity bins except for one, the difference in the
fraction with a \Neiii\ excess is less than 2$\sigma$ significant.
However, for the overall proportion of \Neiii\ excess detections, the Fisher's exact test indicates that there is a probability of less than 0.03 that the incidence of this behavior is the same in the Sy1 and Sy2 samples. Therefore, the full sample in our study supports, with a moderate statistical significance, previous indications that Sy2 host galaxies tend to have higher rates of star formation than Sy1 hosts. However the full sample is heterogeneous and may be subject to a number of biases. For example, it includes a number of radio galaxies, which are known to have low star formation rates -- in fact none of the radio galaxies have \Neiii\ excess. It also contains ULIRGs with Seyfert spectral characteristics, most of which do have \Neiii\ excess. The observed differences depend to some extent on the relative number of galaxies in these two classes. In the following section, we discuss this trend in two relatively complete but smaller subsamples where the biases should be reduced.

\subsection{Results for the 12\micron\ sample and the RSA samples}\label{ss:twosamples}

At this point we study separately the 12\micron\ sample and the RSA sample. Both are complete samples of Seyfert galaxies selected with homogeneous criteria. The former is selected based on their \textit{IRAS} 12\micron\ fluxes whereas the latter is selected based on the optical magnitude of the host galaxy. Given the different selection criteria they they might be affected by star formation in different ways.

\begin{figure*}
\center
\includegraphics[width=0.95\textwidth]{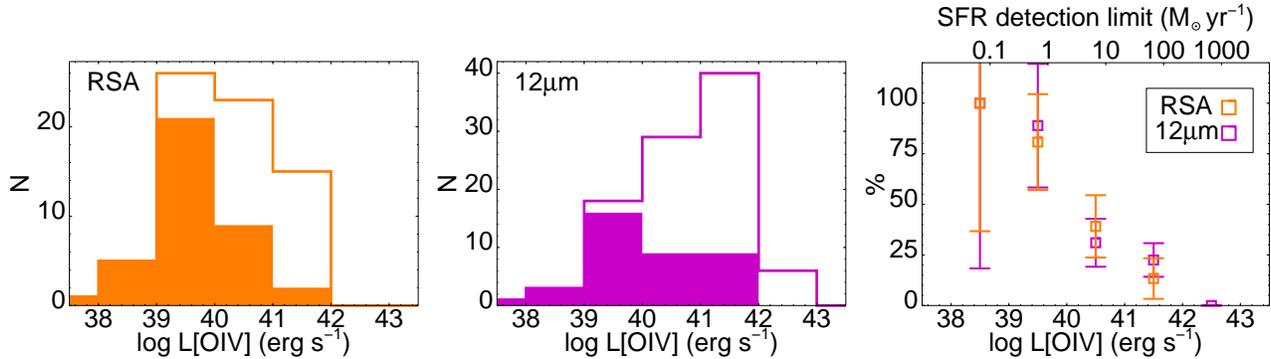}
\caption{Distribution of the \Oiv\ luminosities for the RSA sample (left panel) and 12\micron\ sample (middle panel). Symbols are as in Figure \ref{fig_comparison_sy1sy2_all}.}
\label{fig_comparison_rsa12}
\end{figure*}

As can be seen from Table \ref{tbl_comparison_samples} and Figure \ref{fig_comparison_rsa12}, the fraction of galaxies with a \Neiii\ excess in the RSA sample is higher (by a factor of 1.4) than in the 12\micron\ sample.
The lower number of galaxies with a \Neiii\ excess in the 12\micron\ sample is because the galaxies of this sample are brighter and the sensitivity to star formation of our method is reduced at high \Oiv\ luminosities.
Both effects are clearly seen in Figure \ref{fig_comparison_rsa12}. First, the RSA sample contains a larger fraction of low luminosity Seyferts ($L_{\rm  [O~{\scriptscriptstyle IV}]} \le 10^{40}$\,erg s$^{-1}$) than the 12\micron\ sample. Second, for each AGN luminosity, we are sensitive to a different SFR limit.
In other words, for the brightest Seyferts we are only sensitive to SFR above 100\,$M_{\rm \odot}$ yr$^{-1}$, while for low luminosity AGN we are sensitive to SFR $>0.1\,M_{\rm \odot}$ yr$^{-1}$. 
We note that the fraction of galaxies with star formation in each luminosity bin is similar for both samples (right panel of Figure \ref{fig_comparison_rsa12}).

In both samples the fraction of Seyfert 2s with \Neiii\ excess is slightly larger than that of Seyfert 1s. However due to the small size of the samples the statistical significance of this excess is low. Based on the two sample K-S test, the probability for these differences being due to chance is 0.25 and 0.4 for the 12\micron\ and RSA samples respectively.

\begin{deluxetable}{lcccc}
\tablewidth{0pt}
\tablecaption{Fraction of galaxies with a [\ion{Ne}{3}] excess in the
  RSA and 12\micron\ samples \label{tbl_comparison_samples}}
\tablehead{\colhead{Type} & \colhead{12\micron\ sample} & \colhead{RSA}}
\startdata
Seyfert 1 & 15\slash45 & 12\slash25 \\
Seyfert 2 & 23\slash52 & 25\slash45 \\
Total & 38\slash97 & 37\slash70
\enddata
\tablecomments{In this table we only include galaxies with the \Oiv\ and \Neiii\ emission lines detected. Some galaxies are common to both samples.}
\end{deluxetable}

\section{Line ratios and models}\label{s:models}

We studied the physical conditions in the narrow line region of the AGNs with the photoionization code MAPPINGSIII \citep{Groves2004MAPPINGS}. We used a radiation pressure dominated model that includes the effects of dust. For the input parameters we followed the prescription given by \citet{Groves2006}. Briefly, we assumed a plane parallel geometry and solar abundances. We modeled the input ionizing spectrum with two power-laws with exponential cut-offs \citep{Nagao2001}. We explored the effect of the variations in the total hydrogen column density ranging from $\log$ $n_{\rm H}$ (cm$^{-2}$) = 19.0 to 22.0. This range was chosen since it reproduces the observed line ratios and is in good agreement with the hydrogen column density determined using UV and X-ray observations \citep{Crenshaw2003}. For each hydrogen column density we varied the total pressure ($P/$\kboltz) and the incident ionizing flux ($I_{\rm 0}$). The values for the total pressure are $\log$ $P$/\kboltz\,(K cm$^{-3}$) = 6, 7, 8, 9. These correspond to electron densities, as traced by [\ion{S}{2}]$\lambda$6716\AA\slash$\lambda$6731\AA, of $\sim$ $<$10$^1$, 10$^2$, 10$^3$ and $>$ 10$^4$\,cm$^{-3}$ respectively. The explored range for the ionizing flux is $I_{\rm 0}$ = 0.1 to 0.55. The incident ionizing flux is scaled by the factor 2.416$\times$(($P$/\kboltz)/10$^6$)\,erg cm$^{-2}$ s$^{-1}$ which gives a range in the ionization parameter, $\log$ $U$, from $\sim$ -3 to -2.

\subsection{The [\ion{Ne}{5}] ratio}\label{ss:ne5ratio}

\begin{figure}[h]
\center
\includegraphics[width=0.45\textwidth]{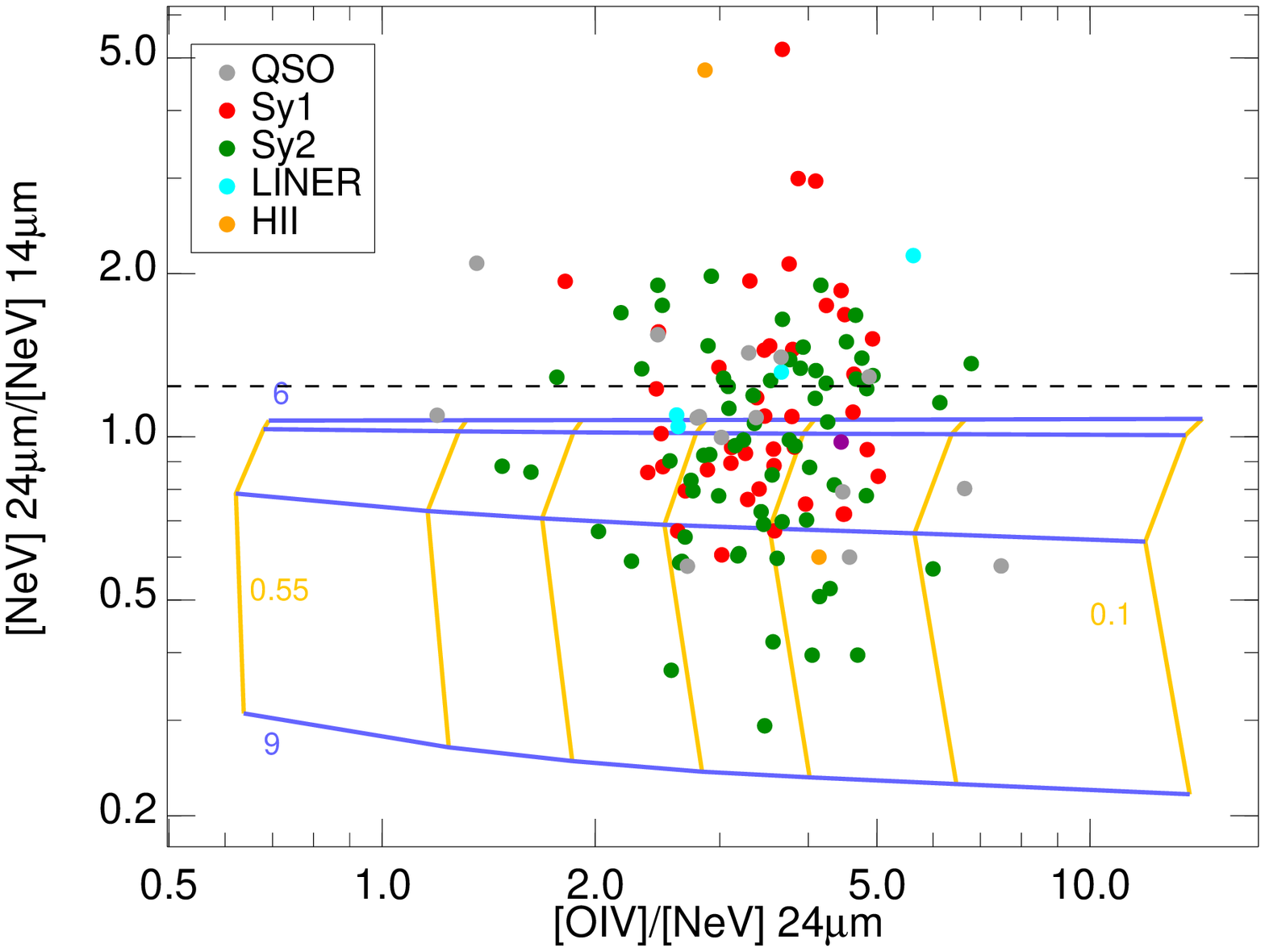}
\caption{Predicted \Nevb\slash\Neva\ ratio vs. \Oiv\slash\Nevb. Galaxy symbols are as in Figure \ref{fig_ne5o4}. The dashed black line indicates the low density limit calculated at 10000 K for the \Nevb\slash\Neva\ ratio (galaxies above this line have densities below the low density limit). The blue lines 
trace constant pressure and the orange lines constant ionization parameter ($I_{\rm 0}$). The input parameters for the model are $\log$ $P$/\kboltz\ = 6, 7, 8, 9 and $I_{\rm 0}$ = 0.1, 0.15, 0.20, 0.25, 0.32, 0.4, 0.55 for constant hydrogen column density $\log$ $n_{\rm H}$ (cm$^{-2}$) = 20.3. The galaxy symbols are as in Figure \ref{fig_ne5o4}.}
\label{fig_hardness}
\end{figure}

Figure \ref{fig_hardness} shows the ratio between the two [\ion{Ne}{5}] emission lines available in the mid-infrared. The critical densities of the \Neva\ and \Nevb\ lines are 3.5$\times$10$^4$\,cm$^{-3}$ and 6.2$\times$10$^3$\,cm$^{-3}$, respectively \citep{Osterbrock2006}, thus they can be used to trace the density of the NLR.
We find however that a large fraction of the galaxies ($\sim$ 30\%) lies above the low density limit ratio. This issue and its possible cause are discussed in detail by \citet{Dudik07} and also addressed by \citet{Tommasin2010} for a sample of Seyfert galaxies.
\citet{Dudik07} concluded that the ratios above the low density limit are due to differential extinction from the obscuring torus, and thus can indicate the inclination angle to our line of sight.
However, similar to \citet{Baum2010}, we do not find any significant difference between the proportion of Seyfert 1 and Seyfert 2 galaxies above the low density limit, as would be predicted by this hypothesis.

We calculated the $A_{\rm 24.3}$\slash $A_{\rm 14.3}$ ratio to quantify the extinction needed to move the points above the low density limit ratio to this value (which would be a lower limit to the extinction since the real ratio may be lower than the low density limit). The ratios are summarized in Table \ref{tbl_extinction} for some infrared extinction laws. As can be seen depending on the extinction law chosen the extinction can increase or decrease the \Nevb\slash\Neva\ ratio.
Since the construction of an extinction law implies some interpolation we also used direct measurements of $A_{\rm 24}$\slash $A_{\rm K}$ and $A_{\rm 15}$\slash $A_{\rm K}$.\footnote{From the extinction laws, the difference between these $A$ values and those at 14.3\micron\ and 24.3\micron\ would be small ($\sim$5\%).}
\citet{Jiang2006} measured the extinction at 15\micron\ and found $A_{\rm 15}$\slash $A_{\rm K}$ $=$ 0.40, although it ranges from 0.25 to 0.55.
For the $A_{\rm 24}$\slash $A_{\rm K}$ ratio the range is 0.28 to 0.65 for different $A_{\rm K}$ bins \citep{Chapman2009_690} and the average value is $\sim$0.5 \citep{Chapman2009_690, Flaherty2007}. 
From these measurements we estimate $A_{\rm 24.3}$\slash $A_{\rm 14.3}$ $=$ 1.2\,$\pm$\,0.3. Again it is close to unity and it is not clear if the extinction would increase or decrease the [\ion{Ne}{5}] ratio.
Moreover the extinction is nearly neutral between the lines with all the extinction laws and therefore there is very little potential effect on the ratio unless the extinction is extremely large.

\tabletypesize{\small}
\begin{deluxetable}{lc}
\tablewidth{0pt}
\tablecaption{Extinction at 24~$\mu{\rm m}$ relative to that at 14~$\mu{\rm m}$\label{tbl_extinction}}
\tablehead{
\colhead{Extinction law} & \colhead{A$_{24}$\slash A$_{14}$}}
\startdata
Rosenthal\tablenotemark{a} & 1.25 \\
McClure 0.3 $<$ $A_{\rm K}$ $<$ 1\tablenotemark{b} & 0.83 \\
McClure 1 $<$ $A_{\rm K}$ $<$ 7\tablenotemark{b} & 0.83 \\
Chiar Galactic Center\tablenotemark{c} & 1.38 \\
Chiar Local ISM\tablenotemark{c} & 0.97
\enddata
\tablenotetext{a}{\citealt{Rosenthal2000}}
\tablenotetext{b}{\citealt{McClure2009}}
\tablenotetext{c}{\citealt{Chiar2006}}
\end{deluxetable}

We also checked if aperture effects can explain the large values of this ratio. The \Neva\ line is observed with the SH module (5\farcs7$\times$11\farcs3), whereas the \Nevb\ is observed with the LH module (11\farcs1$\times$22\farcs3). We would expect a correlation between the \Nevb\slash\Neva\ ratio and the distance if the [\ion{Ne}{5}] emission region was larger than the SH slit. However we do not find this correlation thus we rule out any aperture bias.

Alternatively the atomic parameters for the Ne$^{+4}$ ion may not be sufficiently accurate for the comparison with these observations and the calculated low density limit may be incorrect.

\subsection{Ionization parameter}

In Figure \ref{fig_hardness} we plot the model grid together with the data. In this plot we vary the ionizing flux, $I_{\rm 0}$, and the total pressure, $P/$\kboltz, for a constant total hydrogen column density ($\log$ $n_{\rm H}$\,(cm$^{-2}$) = 20.3). 
We see that the [\ion{Ne}{5}] ratio can be used to measure the pressure. However the pressure (or density) values derived from this ratio have some issues, since a considerable fraction of galaxies are above the low density limit ratio (Section \ref{ss:ne5ratio}).

Also, Figure \ref{fig_hardness} shows that the \Oiv\slash\Nevb\ ratio traces the ionization parameter. The \Oiv\slash\Nevb\ ratio spans a narrow range (less than 1 dex) which is related to the tight correlation found between the \Oiv\ and \Nevb\ luminosities (Section \ref{s:o4ne5}).
This implies a very small range for the ionization parameter (-2.8 $< \log U <$ -2.5).
This range is similar to that found when modeling the optical [\ion{O}{3}] and H$\beta$ emission \citep{Baskin2005, Kraemer1999}.
The critical density of the [\ion{O}{3}] lines is larger than that of these mid-IR lines and for instance the \citet{Baskin2005} models predict electron densities $\log n_{\rm e}$\,(cm$^{-3}$) $=$ 5.8\,$\pm$\,0.7 well above the critical density of the mid-IR lines studied here. Thus it is possible that these lines are produced in different gas clouds.

\subsection{Hydrogen column density}

We use the \Neiii\ and the \Neva\ lines to estimate the total hydrogen (\ion{H}{1} $+$ \ion{H}{2}) column density, $n_{\rm H}$. The \Neva\ line is produced in the inner part of the NLR while the \Neiii\ is produced in a more external region. Thus the \Neiii\slash\Neva\ ratio is subject to change depending on the column density. As expected, the models predict larger \Neiii\slash\Neva\ ratios for larger $n_{\rm H}$.

\begin{figure}[h]
\center
\includegraphics[width=0.45\textwidth]{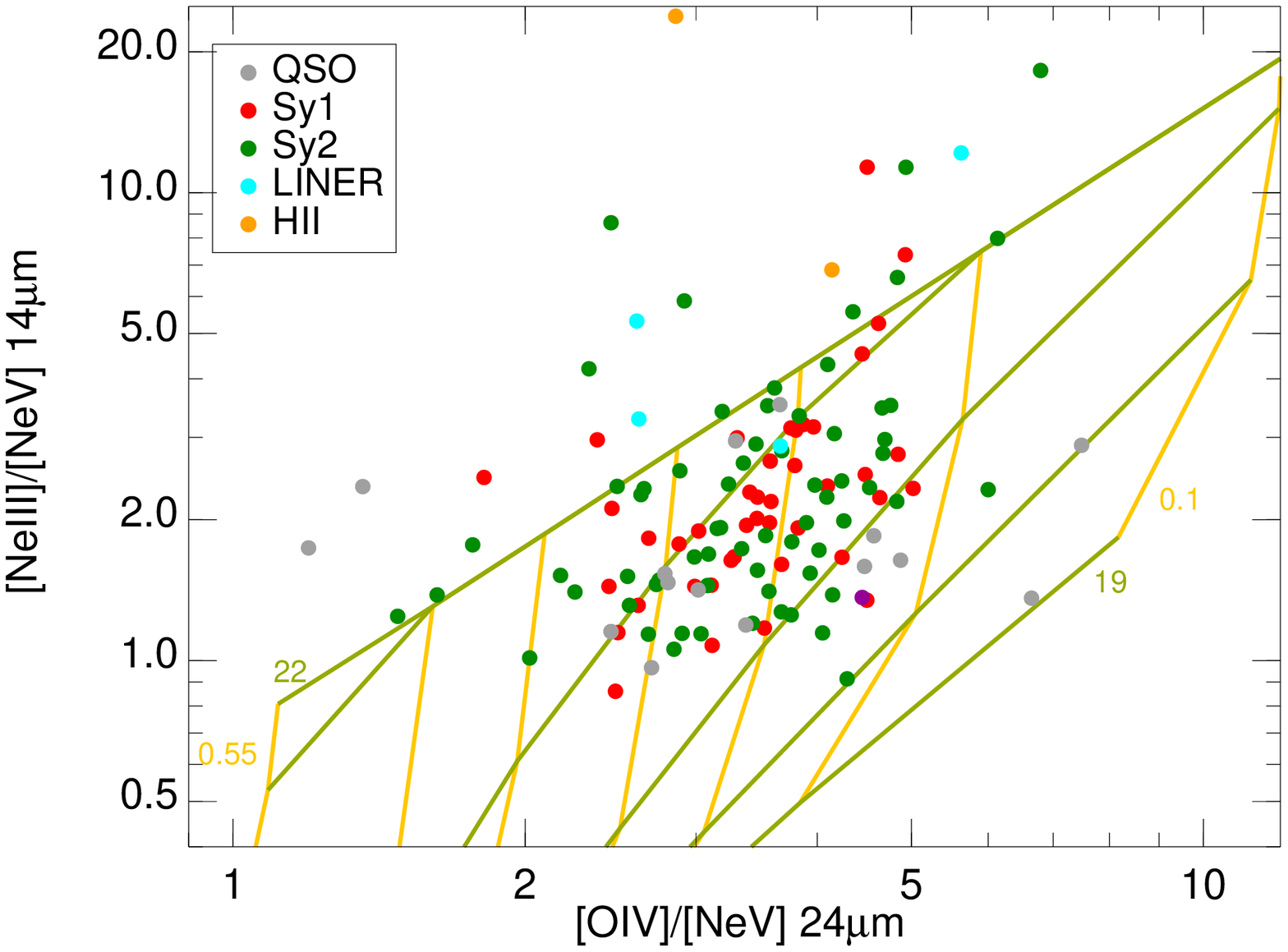}
\caption{Predicted \Neiii\slash\Neva\ ratio vs. \Oiv\slash\Nevb. The green lines trace constant hydrogen column density and the orange lines constant ionization parameter. The input parameters for the model are $\log$ $n_{\rm H}$\,(cm$^{-2}$) = 19.0, 20.0, 20.3, 20.6, 21.0, 22.0 and $I_{\rm 0}$ = 0.1, 0.15, 0.20, 0.25, 0.32, 0.4, 0.55 for constant pressure ($\log$ $P$/\kboltz\ = 8). The galaxy symbols are as in Figure \ref{fig_ne5o4}.}
\label{fig_columndesnity}
\end{figure}

In Figure \ref{fig_columndesnity} we plot the model grid together with the observed ratios. 
The \Neiii\slash\Neva\ ratio is not sensitive to column densities larger than $\log$ $n_{\rm H}$\,(cm$^{-2}$) $>$ 21. We find that the column density for most of the galaxies is in the range 20.3 $<$ $\log$ $n_{\rm H}$\,(cm$^{-2}$) $<$ 21. There is no difference between the predicted hydrogen column density for type 1 and 2 galaxies. This column density range is compatible with the values derived from UV observations although lower than that determined using X-ray data \citep{Crenshaw2003}.
A few galaxies are above the model grid. The \Neiii\slash\Neva\ ratio of these galaxies is larger than that predicted by the models. As we proposed in Sections \ref{s:ne3_agn} and \ref{ss:ne3sf_estimation}, star formation acting in these galaxies is likely producing the extra \Neiii\ emission.

\section{Conclusions}\label{s:conclusions}

We studied a sample of 426 galaxies observed with the \Spitzer/IRS spectrograph in the high resolution mode. Our analysis includes published data for QSO, Seyfert, LINER and \HII\ galaxies as well as unpublished measurements for the RSA Seyfert sample and LIRGs. We explored the relationship between the high-ionization (\Neva, \Nevb\ and \Oiv) and intermediate ionization (\Neiii\ and \Neii) emission lines present in the mid-IR spectra. The median ratios for each class of galaxies are listed in Table \ref{tbl_ratios}.
The main results are as follows:

\begin{enumerate}
\item{There is a tight linear correlation between the fluxes and luminosities of the high-ionization emission lines \Oiv\ and \Nevb. This correlation spans 5 orders of magnitude (38 $<$ $\log$ $L_{\rm [Ne~{\scriptscriptstyle V}]24}$\,(erg s$^{-1}$) $<$ 43), ranging from LINERs, low-luminosity Seyfert galaxies, types 1 and 2, to QSOs.
The typical \Oiv\slash\Nevb\ ratio for AGNs is 3.5, with rms scatter of 0.8. The correlation also holds between the \Oiv\ and the \Neva\ lines, although the scatter is larger. The typical \Oiv\slash\Neva\ ratio for AGNs is 3.4, with rms scatter of 1.4.
}

\item{There is also a good linear correlation between the \Neiii\ and the \Neva\ fluxes and luminosities for Seyfert galaxies and QSO. The \Neiii\slash\Neva\ ratio is 1.9, with rms scatter of 0.8.
Using this correlation and the previous one we calculated the typical \Neiii\slash\Oiv\ ratio for AGNs of 0.6, with rms scatter of 0.3.}

\item{We calculate the \Oiv\ emission due to star formation ($\log L_{\rm [O~{\scriptscriptstyle IV}]}~({\rm erg~s^{-1}})< \log L_{\rm IR}~({\rm erg~s^{-1}}) - 4.7 \pm 0.6
$) and we estimate that it may dominate the total \Oiv\ emission when the intrinsic AGN luminosity is a factor of 20 smaller than the star forming luminosity. In general, we find that the \Oiv\ emission is dominated by the AGN in Seyfert galaxies, whereas star-formation can explain the \Oiv\ emission of optically classified \HII\ galaxies.
}

\item{We do not find any significant difference between the mid-IR high-ionization emission lines in type 1 and in type 2 Seyfert galaxies. Either there are no differences in the conditions in the NLR of the two AGN types, or the effects associated with the different line of sight (i.e., dust extinction) are minimized in the mid-infrared spectra.}

\item{We find that a significant number of Seyfert galaxies from the 12\micron\ sample (30\%) and the RSA sample (40\%) show evidence for excess \Neiii\ emission relative to their \Oiv\ emission associated with star formation. The larger fraction in the RSA sample is explained because the RSA sample contains a larger fraction of low-luminosity AGNs ($L_{\rm [O~{\scriptscriptstyle IV}]}$ $<$ 10$^{41}$\,erg s$^{-1}$) in which these star formation excesses are easier to detect.}

\item{The fraction of Seyfert 2 galaxies with \Neiii\ excess is larger than that of Seyfert 1 with a moderate statistical signiﬁcance in the full sample although this sample may be affected by several bias. However when we consider the 12\micron\ or RSA Seyfert samples separately these differences are not statistically significant due to the smaller size of these samples.}

\item{A considerable fraction (30\%) of the galaxies have \Nevb\slash\Neva\ ratios above the low density limit.
We did not find a connection between the Seyfert type and the ratio of galaxies above the limit.
We are not able to explain this in terms of differential extinction.
}

\item{Our modeling shows that the nebular conditions in the NLRs are remarkably similar among all the AGNs in our sample. This similarity allows us to compare conditions critically in the NLRs of the type 1 and type 2 galaxies. There appear to be no significant overall differences, consistent with the unified model. We constrained the ionization parameter in the range -2.8 $<$ $\log$ $U$ $<$ -2.5 and the hydrogen column density 20 $<$ $\log$ $n_{\rm H}$\,(cm$^{-2}$) $<$ 21.}

\end{enumerate}

The relationships presented in this paper provide an important benchmark for the interpretation of the future mid-IR observations of AGNs and star forming galaxies with JWST/MIRI.

\section*{Acknowledgements}

The authors thank B. Groves and T. D\'iaz-Santos for their help and enlightening discussion. 
We thank the anonymous referee for useful comments and suggestions.
MP-S acknowledges support from the CSIC under grant JAE-Predoc-2007. MP-S also thanks the Steward Observatory for their hospitality during his stay while part of this work was done.
AA-H and MP-S acknowledge support from the Spanish Plan Nacional del Espacio under grant ESP2007-65475-C02-01. This work was partially supported by Caltech/JPL through contract 1255094 to University of Arizona. AA-H also acknowledges support for this work from the Spanish Ministry of Science and Innovation through Proyecto Intramural Especial under grant number 200850I003 and from Plan Nacional de Astronomia y Astrofisica under grant number AYA2009-05705-E.
This research has made use of the NASA/IPAC Extragalactic Database (NED) which is operated by the Jet Propulsion Laboratory, California Institute of Technology, under contract with the National Aeronautics and Space Administration. This research also used the VizieR catalogue Service \citep{Ochsenbein2000Vizier}.

\begin{appendix}
\section{Calculating the AGN contribution}\label{apxContribution}

In this appendix we explain briefly the method  used in Sections \ref{s:ne3sf} and \ref{s:ne3sf_sy1sy2} to calculate the star formation fraction of the \Neiii\ emission.

The total \Neiii\ emission includes two components, one from the AGN and other from star formation (SF).
\begin{equation}
\rm [Ne~{\scriptstyle III}] = \rm [Ne~{\scriptstyle III}]_{SF} + \rm [Ne~{\scriptstyle III}]_{AGN}
\end{equation}

If we assume that all the \Oiv\ is produced by the AGN we can use the typical 
\Neiii\slash\Oiv\ ratio observed in
Seyfert galaxies with low star formation (Figure \ref{fig_lo4_o4ne3}) to estimate the amount of \Neiii\ emission coming from the AGN.
\begin{equation}
\rm [Ne~{\scriptstyle III}] = \rm [Ne~{\scriptstyle III}]_{SF} + \rm [O~{\scriptstyle IV}]\times \left( \frac{\rm [Ne~{\scriptstyle III}]}{[\rm O~{\scriptstyle IV}]} \right)_{AGN}
\end{equation}

Finally we obtain the fraction of \Neiii\ from star formation.
\begin{equation}
\frac{\rm [Ne~{\scriptstyle III}]_{SF}}{\rm [Ne~{\scriptstyle III}]} = 1 - \frac{\rm [O~{\scriptstyle IV}]}{\rm [Ne~{\scriptstyle III}]} \times \left( \frac{\rm [Ne~{\scriptstyle III}]}{\rm [O~{\scriptstyle IV}]} \right)_{AGN}
\end{equation}

This method can be applied as well to other line ratios. The only assumption is that one of the lines in the ratio is uniquely produced by the AGN while star formation and the AGN contributes to the other line (e.g. the \Neiii\ and [\ion{Ne}{5}] lines).
Also note that this method makes use of a typical AGN ratio thus the estimated star formation contribution to an emission line for a single object will be uncertain and dependent on the ionization paremeter.

\end{appendix}

\clearpage
\LongTables
\setcounter{table}{0}

\tabletypesize{\scriptsize}
\begin{deluxetable*}{lcccccccccc}
\tablewidth{0pt}
\tablecaption{Sample and Mid-Infrared Emission Line Fluxes\label{tbl_fluxes_sample}}
\tablehead{ & \colhead{R.A.\tablenotemark{a}} & \colhead{Decl.\tablenotemark{a}} & \colhead{Dist.\tablenotemark{b}} & \colhead{Spect.}
& \colhead{[\ion{Ne}{2}]} &  \colhead{[\ion{Ne}{5}]} & \colhead{[\ion{Ne}{3}]} &\colhead{[\ion{Ne}{5}]} & \colhead{[\ion{O}{4}]} & \\
\colhead{Name} & \colhead{(J2000.0)} & \colhead{(J2000.0)} & \colhead{(Mpc)} & \colhead{Class.\tablenotemark{a}} & \colhead{12.81\micron} & \colhead{14.32\micron} & \colhead{15.56\micron}  & \colhead{24.32\micron} & \colhead{25.89\micron} & \colhead{Ref.}}

\startdata
Mrk~335 $^d$ & 00 06 19.5 & +20 12 10 & 113 & Sy1 & 0.25 & 0.38 & 0.61 & 2.0 & 7.2 & 4 \\
NGC~23 & 00 09 53.4 & +25 55 25 & 64.5 & \ion{H}{2} & 96 & $<$0.50 & 13 & $<$0.44 & 1.4 & 2 \\
NGC~24 & 00 09 56.5 & --24 57 47 & 7.30 & \nodata & 3.1 & $<$0.88 & 1.3 & \nodata & $<$0.42 & 5 \\
Mrk~938 $^d$ & 00 11 06.5 & --12 06 26 & 85.3 & Sy2 & 52 & $<$2.2 & 6.4 & $<$0.37 & $<$0.66 & 4 \\
IRAS~F00188--0856 & 00 21 26.5 & --08 39 26 & 600 & LINER & 4.7 & $<$0.18 & 0.69 & $<$1.6 & $<$0.90 & 3 \\
IRAS~00198--7926 $^d$ & 00 21 53.6 & --79 10 07 & 329 & Sy2 & 6.2 & 12 & 14 & 11 & 33 & 4 \\
PG~0026+129 & 00 29 13.6 & +13 16 03 & 672 & QSO & 0.23 & 0.47 & 0.79 & $<$0.33 & 2.1 & 1 \\
ESO~012-G021 $^d$ & 00 40 46.2 & --79 14 24 & 144 & Sy1 & 12 & 3.2 & 6.4 & 4.6 & 16 & 4 \\
IRAS~F00397--1312 & 00 42 15.5 & --12 56 02 & 1329 & \ion{H}{2} & 4.4 & $<$0.20 & 2.7 & $<$1.5 & $<$1.2 & 3 \\
NGC~253 & 00 47 33.1 & --25 17 17 & 2.50 & \ion{H}{2} & 2832 & $<$21 & 205 & $<$73 & 155 & 8 \\
Mrk~348 $^d$ & 00 48 47.1 & +31 57 25 & 65.1 & Sy2 & 16 & 5.8 & 20 & 4.9 & 18 & 4 \\
NGC~278 & 00 52 04.3 & +47 33 01 & 11.4 & \nodata & 18 & $<$0.070 & 3.0 & $<$0.59 & $<$0.62 & 7 \\
PG~0050+124 & 00 53 34.9 & +12 41 36 & 273 & QSO & 1.9 & 5.5 & 4.5 & $<$2.1 & 2.7 & 1 \\
IRAS~00521--7054 $^d$ & 00 53 56.1 & --70 38 04 & 310 & Sy2 & 5.8 & 5.8 & 8.1 & 2.4 & 8.6 & 4 \\
MCG~+12-02-001 & 00 54 03.6 & +73 05 11 & 68.1 & \ion{H}{2} & 242 & $<$1.00 & 43 & $<$1.4 & 3.8 & 2 \\
UGC~556 & 00 54 50.3 & +29 14 47 & 66.1 & \ion{H}{2} & 40 & $<$0.21 & 7.1 & $<$0.73 & $<$1.6 & 2 \\
NGC~337 & 00 59 50.1 & --07 34 40 & 22.4 & \ion{H}{2} & 19 & $<$0.74 & 8.0 & \nodata & $<$0.49 & 5 \\
ESO~541-IG012 $^d$ & 01 02 17.5 & --19 40 08 & 253 & Sy2 & 1.9 & 2.2 & 2.0 & 1.2 & 5.0 & 4 \\
IRAS~01003--2238 & 01 02 50.0 & --22 21 57 & 550 & \ion{H}{2} & 3.1 & $<$0.30 & 1.3 & $<$0.30 & $<$0.30 & 3 \\
3C~33 & 01 08 52.9 & +13 20 13 & 269 & Sy2 & 3.9 & 2.0 & 5.3 & \nodata & 8.1 & 6 \\
NGC~404 & 01 09 27.0 & +35 43 04 & 3.24 & LINER & 3.1 & $<$0.16 & 1.7 & $<$0.31 & 0.90 & 2 \\
NGC~424 $^d$ & 01 11 27.6 & --38 05 00 & 50.8 & Sy2 & 8.7 & 16 & 18 & 6.4 & 26 & 4 \\
NGC~526A $^d$ & 01 23 54.4 & --35 03 55 & 83.0 & Sy1 & 5.8 & 6.3 & 10 & 5.9 & 19 & 4 \\
NGC~513 $^d$ & 01 24 26.9 & +33 47 58 & 85.0 & Sy2 & 13 & 1.9 & 4.4 & 1.1 & 6.5 & 4 \\
NGC~520 & 01 24 35.1 & +03 47 32 & 30.2 & \ion{H}{2} & 45 & $<$0.64 & 7.5 & $<$1.4 & 8.1 & 8 \\
NGC~584 & 01 31 20.8 & --06 52 05 & 20.0 & \nodata & \nodata & $<$0.92 & 1.7 & \nodata & $<$0.30 & 5 \\
NGC~613 & 01 34 18.2 & --29 25 06 & 15.0 & \ion{H}{2} & 131 & 0.67 & 16 & 3.2 & 9.1 & 7 \\
NGC~633 & 01 36 23.4 & --37 19 17 & 74.2 & \ion{H}{2} & 50 & $<$0.50 & 6.1 & $<$0.59 & 1.9 & 2 \\
NGC~628 & 01 36 41.8 & +15 47 00 & 7.30 & \nodata & 6.2 & $<$1.00 & \nodata & \nodata & $<$0.38 & 5 \\
IRAS~01364--1042 & 01 38 52.9 & --10 27 11 & 215 & \ion{H}{2} & 7.9 & $<$0.32 & 1.5 & $<$0.60 & $<$0.91 & 2 \\
NGC~660 & 01 43 02.4 & +13 38 42 & 12.1 & \ion{H}{2} & 346 & $<$0.57 & 37 & $<$4.5 & 28 & 2 \\
Mrk~573 & 01 43 57.8 & +02 20 59 & 73.8 & Sy2 & \nodata & 18 & 24 & \nodata & $<$79 & 6 \\
III~Zw~035 & 01 44 30.5 & +17 06 05 & 120 & \nodata & 3.8 & $<$0.17 & 0.19 & $<$1.9 & $<$1.3 & 2 \\
IRAS~01475--0740 $^d$ & 01 50 02.7 & --07 25 48 & 76.7 & Sy2 & 14 & 6.4 & 9.9 & 1.9 & 6.5 & 4 \\
3C~55 & 01 57 10.5 & +28 51 37 & 4522 & Sy2 & \nodata & 1.1 & 2.0 & \nodata & \nodata & 6 \\
Mrk~1014 & 01 59 50.2 & +00 23 40 & 781 & Sy1 & 6.6 & 7.4 & 9.7 & 5.0 & 13 & 3 \\
NGC~788 $^c$ & 02 01 06.4 & --06 48 55 & 58.9 & Sy2 & 6.1 & 5.3 & 14 & 7.9 & 23 & 2 \\
NGC~855 & 02 14 03.6 & +27 52 37 & 9.60 & \nodata & 5.5 & $<$0.72 & \nodata & \nodata & $<$0.40 & 5 \\
NGC~891 & 02 22 33.4 & +42 20 56 & 8.60 & \ion{H}{2} & 8.6 & $<$0.040 & 0.84 & $<$0.62 & $<$1.0 & 7 \\
Mrk~1034~NED01 $^d$ & 02 23 18.9 & +32 11 18 & 148 & Sy1 & 19 & $<$1.1 & 1.5 & $<$0.59 & $<$0.77 & 4 \\
Mrk~1034~NED02 $^d$ & 02 23 22.0 & +32 11 49 & 148 & Sy1 & 35 & 1.1 & 3.6 & $<$0.60 & 2.7 & 4 \\
UGC~1845 & 02 24 08.0 & +47 58 11 & 66.7 & \ion{H}{2} & 106 & $<$0.59 & 11 & $<$0.82 & 3.1 & 2 \\
ESO~545-G013 $^d$ & 02 24 40.6 & --19 08 31 & 148 & Sy1 & 10 & 4.8 & 11 & 3.2 & 12 & 4 \\
NGC~925 & 02 27 16.9 & +33 34 45 & 9.10 & \ion{H}{2} & 10 & $<$1.0 & 5.1 & \nodata & $<$0.36 & 5 \\
NGC~931 $^d$ & 02 28 14.5 & +31 18 41 & 72.2 & Sy1 & 5.5 & 14 & 15 & 14 & 43 & 4 \\
NGC~1055 & 02 41 45.2 & +00 26 35 & 11.3 & Sy2 & 26 & $<$0.19 & 2.6 & $<$0.58 & 1.2 & 7 \\
NGC~1068 $^c$ $^d$ & 02 42 40.7 & --00 00 47 & 16.3 & Sy2 & 461 & 895 & 1355 & 808 & 2061 & 2 \\
NGC~1056 $^d$ & 02 42 48.3 & +28 34 27 & 22.1 & Sy2 & 34 & $<$1.8 & 10 & $<$1.2 & 1.4 & 4 \\
NGC~1058 $^c$ & 02 43 30.0 & +37 20 28 & 7.42 & Sy2 & 1.0 & $<$0.46 & \nodata & $<$0.33 & $<$0.50 & 2 \\
UGC~2238 & 02 46 17.5 & +13 05 44 & 94.5 & \ion{H}{2} & 65 & $<$0.47 & 10 & $<$2.6 & 5.5 & 2 \\
NGC~1097 $^c$ $^d$ & 02 46 19.1 & --30 16 29 & 17.7 & Sy1 & 165 & $<$0.74 & 20 & $<$2.4 & 5.0 & 2 \\
IRAS~02438+2122 & 02 46 39.1 & +21 35 10 & 102 & LINER & 18 & $<$0.33 & 1.6 & $<$2.2 & $<$2.6 & 2 \\
NGC~1125 $^d$ & 02 51 40.3 & --16 39 03 & 47.2 & Sy2 & 16 & 5.1 & 16 & 9.7 & 40 & 4 \\
NGC~1142 $^d$ & 02 55 12.2 & --00 11 00 & 126 & Sy2 & 17 & 0.92 & 5.4 & 1.8 & 5.3 & 4 \\
Mrk~1066 & 02 59 58.6 & +36 49 14 & 51.9 & Sy2 & \nodata & 18 & 52 & \nodata & \nodata & 6 \\
MCG~--02-08-039 $^d$ & 03 00 30.6 & --11 24 56 & 131 & Sy2 & 3.9 & 6.6 & 9.8 & 5.2 & 14 & 4 \\
NGC~1194 $^d$ & 03 03 49.1 & --01 06 13 & 58.9 & Sy2 & 3.8 & 4.3 & 7.4 & 3.8 & 15 & 4 \\
NGC~1204 & 03 04 39.9 & --12 20 28 & 64.8 & \ion{H}{2} & 54 & $<$0.34 & 5.0 & $<$2.0 & $<$1.6 & 2 \\
NGC~1222 & 03 08 56.7 & --02 57 18 & 32.3 & \ion{H}{2} & 81 & $<$0.57 & 89 & $<$0.78 & 9.9 & 8 \\
3C~79 & 03 10 00.1 & +17 05 58 & 1295 & Sy2 & \nodata & 0.11 & 0.20 & \nodata & 0.70 & 6 \\
NGC~1241 $^c$ $^d$ & 03 11 14.6 & --08 55 19 & 58.5 & Sy2 & 14 & 2.5 & 8.4 & 1.5 & 4.8 & 2 \\
NGC~1266 & 03 16 00.8 & --02 25 38 & 30.0 & LINER & 29 & $<$0.92 & 9.7 & \nodata & $<$1.4 & 5 \\
NGC~1291 & 03 17 18.6 & --41 06 29 & 10.8 & \ion{H}{2} & 5.2 & $<$0.86 & 7.2 & \nodata & 1.4 & 5 \\
IRAS~F03158+4227 & 03 19 12.4 & +42 38 28 & 631 & \nodata & 5.8 & $<$1.1 & 0.94 & $<$1.4 & $<$1.8 & 3 \\
NGC~1275 $^c$ & 03 19 48.2 & +41 30 42 & 76.2 & Sy2 & 43 & $<$1.2 & 21 & $<$2.5 & 9.6 & 2 \\
NGC~1316 & 03 22 41.7 & --37 12 29 & 24.3 & LINER & 13 & $<$0.61 & 11 & \nodata & 2.6 & 5 \\
NGC~1320 $^d$ & 03 24 48.7 & --03 02 32 & 38.3 & Sy2 & 9.6 & 11 & 14 & 7.5 & 27 & 4 \\
NGC~1365 $^c$ $^d$ & 03 33 36.4 & --36 08 25 & 23.9 & Sy1 & 156 & 19 & 61 & 40 & 151 & 2 \\
NGC~1358 $^c$ & 03 33 39.7 & --05 05 21 & 58.2 & Sy2 & 5.1 & 3.5 & 8.2 & $<$1.6 & 9.1 & 2 \\
NGC~1377 & 03 36 39.1 & --20 54 08 & 25.2 & \ion{H}{2} & 4.1 & $<$1.9 & 2.6 & \nodata & $<$1.7 & 5 \\
NGC~1386 $^c$ $^d$ & 03 36 46.2 & --35 59 57 & 12.5 & Sy2 & 14 & 37 & 39 & 34 & 97 & 2 \\
NGC~1404 & 03 38 51.9 & --35 35 39 & 18.5 & \nodata & 1.4 & $<$0.97 & 1.00 & \nodata & $<$0.35 & 5 \\
NGC~1433 $^c$ $^d$ & 03 42 01.6 & --47 13 19 & 15.4 & Sy2 & \nodata & $<$1.6 & \nodata & $<$1.9 & $<$6.9 & 2 \\
NGC~1448 & 03 44 31.9 & --44 38 41 & 11.5 & \nodata & 8.2 & 3.7 & 5.0 & 3.6 & 16 & 7 \\
IC~342 & 03 46 48.5 & +68 05 46 & 4.60 & \ion{H}{2} & 615 & $<$2.4 & 37 & $<$4.9 & $<$7.7 & 8 \\
IRAS~03450+0055 $^d$ & 03 47 40.2 & +01 05 14 & 136 & Sy1 & 1.1 & $<$1.5 & 1.8 & $<$1.9 & 2.5 & 4 \\
NGC~1482 & 03 54 38.9 & --20 30 08 & 23.2 & \ion{H}{2} & 457 & $<$2.6 & 56 & \nodata & $<$6.4 & 5 \\
IRAS~03521+0028 & 03 54 42.2 & +00 37 03 & 724 & LINER & 2.8 & $<$0.45 & 1.3 & $<$0.48 & $<$0.90 & 3 \\
NGC~1512 & 04 03 54.3 & --43 20 55 & 11.8 & \ion{H}{2} & 31 & $<$1.1 & 4.7 & \nodata & $<$0.83 & 5 \\
3C~109 & 04 13 40.4 & +11 12 13 & 1589 & Sy1 & 0.030 & 0.070 & 0.11 & \nodata & 0.27 & 6 \\
IC~2056 & 04 16 24.5 & --60 12 24 & 13.8 & \nodata & 13 & $<$0.16 & 2.2 & $<$0.10 & 0.75 & 7 \\
NGC~1559 & 04 17 35.8 & --62 47 01 & 12.7 & \nodata & 15 & $<$0.44 & 2.3 & $<$0.36 & $<$0.69 & 7 \\
NGC~1566 $^c$ $^d$ & 04 20 00.4 & --54 56 16 & 21.3 & Sy1 & 15 & 1.1 & 9.6 & $<$1.9 & 8.4 & 2 \\
NGC~1569 & 04 30 49.1 & +64 50 52 & 4.60 & Sy1 & 19 & $<$0.38 & 188 & $<$1.4 & 29 & 7 \\
3C~120 $^d$ & 04 33 11.1 & +05 21 15 & 145 & Sy1 & 7.8 & 17 & 28 & 29 & 123 & 4 \\
Mrk~618 $^d$ & 04 36 22.2 & --10 22 33 & 156 & Sy1 & 16 & 3.9 & 5.4 & $<$1.3 & 10 & 4 \\
IRAS~F04385--0828 $^d$ & 04 40 55.0 & --08 22 22 & 65.4 & Sy2 & 14 & 2.3 & 7.1 & $<$1.4 & 8.6 & 4 \\
NGC~1667 $^c$ $^d$ & 04 48 37.1 & --06 19 11 & 65.9 & Sy2 & 10.0 & $<$0.32 & 6.9 & 1.7 & 6.3 & 2 \\
NGC~1705 & 04 54 13.5 & --53 21 39 & 5.10 & \ion{H}{2} & 3.2 & $<$0.91 & 12 & \nodata & 2.5 & 5 \\
ESO~033-G002 $^d$ & 04 55 59.0 & --75 32 28 & 78.6 & Sy2 & 2.1 & 6.3 & 9.2 & 5.3 & 14 & 4 \\
NGC~1792 & 05 05 14.4 & --37 58 50 & 12.5 & \ion{H}{2} & 23 & 0.41 & 2.1 & $<$0.25 & 0.96 & 7 \\
NGC~1808 & 05 07 42.3 & --37 30 47 & 12.6 & Sy2 & 177 & $<$0.91 & 17 & $<$8.4 & $<$9.5 & 7 \\
ESO~362-G018 $^d$ & 05 19 35.8 & --32 39 27 & 53.8 & Sy1 & 12 & 3.3 & 7.5 & 2.6 & 8.9 & 4 \\
IRAS~F05189--2524 & 05 21 01.5 & --25 21 45 & 190 & Sy2 & 21 & 18 & 18 & 12 & 24 & 3 \\
IRAS~05187--1017 & 05 21 06.5 & --10 14 46 & 126 & LINER & 10 & $<$0.15 & 1.9 & $<$0.66 & $<$0.92 & 2 \\
ESO~253-G003 $^d$ & 05 25 18.1 & --46 00 21 & 188 & Sy2 & 16 & 6.8 & 18 & 7.2 & 24 & 4 \\
UGC~3351 & 05 45 47.9 & +58 42 03 & 64.1 & \ion{H}{2} & 67 & $<$0.50 & 9.6 & $<$0.31 & 0.83 & 2 \\
UGCA~116 & 05 55 42.6 & +03 23 31 & 11.6 & \ion{H}{2} & 6.0 & $<$0.28 & 117 & $<$1.1 & 8.5 & 7 \\
IRAS~F05563--3820 $^d$ & 05 58 02.0 & --38 20 04 & 149 & Sy1 & 3.9 & 2.5 & 4.8 & $<$0.79 & 5.3 & 4 \\
IRAS~F06035--7102 & 06 02 54.0 & --71 03 10 & 358 & \ion{H}{2} & 7.0 & $<$0.48 & 1.7 & $<$0.81 & $<$3.0 & 3 \\
ESO~121-G006 & 06 07 29.9 & --61 48 27 & 14.5 & \nodata & 16 & 0.65 & 3.2 & $<$0.59 & 4.6 & 7 \\
Mrk~3 & 06 15 36.4 & +71 02 15 & 60.6 & Sy2 & 100 & 64 & 179 & \nodata & 214 & 6 \\
NGC~2146 & 06 18 37.7 & +78 21 25 & 16.5 & \ion{H}{2} & 625 & $<$2.8 & 91 & $<$3.7 & 19 & 8 \\
UGCA~127 & 06 20 55.7 & --08 29 44 & 10.2 & \nodata & 9.3 & $<$0.15 & 1.5 & $<$0.97 & 3.4 & 7 \\
IRAS~F06206--6315 & 06 21 01.2 & --63 17 23 & 421 & Sy2 & 6.6 & 2.3 & 2.9 & 2.0 & 3.0 & 3 \\
NGC~2273 $^c$ & 06 50 08.7 & +60 50 44 & 26.5 & Sy2 & 40 & 4.5 & 19 & 6.1 & 14 & 2 \\
Mrk~6 $^d$ & 06 52 12.3 & +74 25 37 & 81.7 & Sy1 & 28 & 9.4 & 49 & 10 & 48 & 4 \\
IRAS~07145--2914 & 07 16 31.2 & --29 19 28 & 25.8 & Sy2 & \nodata & 83 & 168 & \nodata & \nodata & 6 \\
NGC~2369 & 07 16 37.7 & --62 20 37 & 46.2 & \ion{H}{2} & 124 & $<$0.51 & 10 & $<$0.91 & 1.8 & 2 \\
NGC~2388 & 07 28 53.4 & +33 49 08 & 58.9 & \ion{H}{2} & 121 & $<$1.2 & 9.8 & $<$0.79 & 1.4 & 2 \\
MCG~+02-20-003 & 07 35 43.4 & +11 42 33 & 71.1 & \ion{H}{2} & 60 & $<$0.47 & 6.3 & $<$0.53 & $<$1.2 & 2 \\
NGC~2403 & 07 36 51.4 & +65 36 09 & 3.20 & \ion{H}{2} & 5.1 & $<$0.71 & 2.1 & \nodata & $<$0.36 & 5 \\
Mrk~9 $^d$ & 07 36 57.0 & +58 46 13 & 176 & Sy1 & 3.2 & 2.2 & 1.9 & 2.2 & 5.6 & 4 \\
3C~184 & 07 39 24.5 & +70 23 10 & 6559 & Sy2 & 0.19 & 0.17 & 0.65 & \nodata & \nodata & 6 \\
Mrk~79 $^d$ & 07 42 32.8 & +49 48 34 & 96.7 & Sy1 & 10 & 6.6 & 20 & 13 & 42 & 4 \\
ESO~209-G009 & 07 58 15.4 & --49 51 15 & 11.8 & \nodata & 9.8 & $<$0.26 & 1.4 & $<$0.12 & $<$0.29 & 7 \\
IRAS~F07598+6508 & 08 04 33.1 & +64 59 48 & 703 & \nodata & 3.9 & $<$0.75 & 2.5 & $<$3.0 & $<$1.8 & 3 \\
3C~196 & 08 13 36.0 & +48 13 02 & 5572 & Sy1 & \nodata & 0.30 & 0.46 & \nodata & \nodata & 6 \\
IRAS~08311--2459 & 08 33 20.6 & --25 09 33 & 460 & Sy1 & 24 & 12 & 23 & 9.8 & 26 & 3 \\
UGC~4459 & 08 34 07.2 & +66 10 54 & 2.78 & \nodata & \nodata & $<$0.89 & 0.98 & \nodata & $<$0.37 & 5 \\
NGC~2623 & 08 38 24.1 & +25 45 16 & 77.4 & Sy2 & 56 & 2.7 & 15 & 2.2 & 9.6 & 8 \\
NGC~2639 $^c$ $^d$ & 08 43 38.1 & +50 12 20 & 44.6 & Sy1 & 8.9 & $<$0.45 & 4.9 & $<$0.82 & 1.8 & 2 \\
PG~0838+770 & 08 44 45.3 & +76 53 09 & 615 & QSO & 0.41 & 0.32 & 0.56 & $<$0.19 & 1.3 & 1 \\
PG~0844+349 & 08 47 42.5 & +34 45 04 & 287 & QSO & 0.42 & 0.30 & 1.0 & 0.42 & 1.5 & 1 \\
NGC~2681 & 08 53 32.7 & +51 18 49 & 12.5 & \nodata & 6.9 & $<$0.15 & 3.4 & $<$0.36 & 2.2 & 7 \\
NGC~2685 $^c$ & 08 55 34.7 & +58 44 03 & 12.7 & Sy2 & \nodata & $<$0.28 & \nodata & $<$0.22 & 0.43 & 2 \\
NGC~2655 $^c$ $^d$ & 08 55 37.7 & +78 13 23 & 20.2 & Sy2 & 13 & $<$1.3 & 7.2 & $<$1.3 & $<$8.0 & 2 \\
IRAS~F08572+3915 & 09 00 25.4 & +39 03 54 & 259 & \nodata & 7.2 & $<$0.75 & 2.0 & $<$5.4 & $<$6.0 & 3 \\
IRAS~09022--3615 & 09 04 12.7 & --36 27 01 & 269 & \nodata & 57 & $<$2.4 & 40 & $<$3.6 & 6.7 & 3 \\
NGC~2798 & 09 17 22.9 & +41 59 59 & 26.2 & \ion{H}{2} & 215 & $<$1.9 & 33 & \nodata & $<$5.1 & 5 \\
Mrk~704 $^d$ & 09 18 26.0 & +16 18 19 & 128 & Sy1 & \nodata & 3.9 & 5.6 & $<$3.7 & 12 & 4 \\
NGC~2841 & 09 22 02.6 & +50 58 35 & 14.1 & LINER & 5.7 & $<$4.5 & 6.7 & \nodata & 0.78 & 5 \\
NGC~2915 & 09 26 11.5 & --76 37 34 & 3.80 & \ion{H}{2} & 3.2 & $<$0.75 & 11 & \nodata & 0.45 & 5 \\
NGC~2903 & 09 32 10.1 & +21 30 03 & 8.30 & \ion{H}{2} & 182 & $<$0.82 & 14 & $<$0.94 & $<$1.7 & 7 \\
UGC~5101 & 09 35 51.7 & +61 21 11 & 172 & LINER & 34 & 2.6 & 14 & 2.8 & 7.3 & 3 \\
NGC~2992 $^c$ $^d$ & 09 45 42.1 & --14 19 34 & 33.2 & Sy2 & 46 & 22 & 61 & 28 & 131 & 2 \\
NGC~2976 & 09 47 15.5 & +67 54 59 & 3.60 & \ion{H}{2} & 7.4 & $<$0.99 & 2.8 & \nodata & $<$0.31 & 5 \\
NGC~3059 & 09 50 08.2 & --73 55 19 & 14.2 & \nodata & 27 & $<$0.18 & 3.1 & $<$0.69 & $<$1.5 & 7 \\
Mrk~1239 $^d$ & 09 52 19.1 & --01 36 43 & 86.7 & Sy1 & 9.4 & 3.4 & 9.4 & 3.2 & 16 & 4 \\
NGC~3049 & 09 54 49.6 & +09 16 15 & 23.9 & \ion{H}{2} & 39 & $<$0.89 & 6.0 & \nodata & $<$1.00 & 5 \\
NGC~3031 $^c$ $^d$ & 09 55 33.2 & +69 03 55 & 3.64 & Sy1 & 27 & $<$0.50 & 21 & $<$1.4 & 4.5 & 2 \\
M~82 & 09 55 52.7 & +69 40 45 & 3.60 & \ion{H}{2} & 506 & $<$0.89 & 81 & $<$15 & 46 & 7 \\
Holmberg~IX & 09 57 32.0 & +69 02 45 & 2.64 & \nodata & \nodata & $<$0.64 & \nodata & \nodata & $<$0.38 & 5 \\
NGC~3081 $^c$ & 09 59 29.5 & --22 49 34 & 34.3 & Sy2 & 12 & 30 & 34 & 39 & 118 & 2 \\
3C~234 $^d$ & 10 01 49.5 & +28 47 08 & 897 & Sy2 & \nodata & 2.3 & 3.4 & 2.9 & 9.0 & 4 \\
NGC~3079 $^c$ $^d$ & 10 01 57.8 & +55 40 47 & 16.2 & Sy2 & 108 & 0.91 & 24 & $<$0.89 & 11 & 2 \\
NGC~3110 & 10 04 02.1 & --06 28 29 & 72.4 & \ion{H}{2} & 75 & $<$0.68 & 9.9 & $<$0.62 & 1.2 & 2 \\
PG~1001+054 & 10 04 20.1 & +05 13 00 & 771 & QSO & 0.40 & $<$0.21 & 0.21 & $<$0.30 & 0.52 & 1 \\
NGC~3175 & 10 14 42.1 & --28 52 19 & 13.4 & \ion{H}{2} & 34 & $<$0.35 & 2.5 & $<$0.86 & 1.1 & 7 \\
IC~2560 $^c$ & 10 16 18.7 & --33 33 49 & 41.7 & Sy2 & 16 & 19 & 36 & 18 & 56 & 2 \\
NGC~3147 $^c$ $^d$ & 10 16 53.7 & +73 24 02 & 40.3 & Sy2 & 4.5 & $<$0.59 & 3.4 & $<$0.91 & $<$2.5 & 2 \\
NGC~3185 $^c$ & 10 17 38.6 & +21 41 17 & 17.7 & Sy2 & 9.6 & 1.0 & 3.6 & 1.8 & 8.2 & 2 \\
NGC~3190 & 10 18 05.6 & +21 49 55 & 20.9 & LINER & 8.3 & $<$0.67 & 4.8 & \nodata & 0.93 & 5 \\
NGC~3184 & 10 18 17.0 & +41 25 27 & 11.1 & \ion{H}{2} & 18 & $<$0.83 & 2.2 & \nodata & $<$0.29 & 5 \\
NGC~3198 & 10 19 54.9 & +45 32 59 & 13.7 & \nodata & 15 & $<$1.00 & 0.81 & \nodata & $<$0.32 & 5 \\
NGC~3227 $^c$ $^d$ & 10 23 30.6 & +19 51 54 & 15.7 & Sy1 & 66 & 23 & 74 & 18 & 70 & 2 \\
NGC~3245 & 10 27 18.4 & +28 30 26 & 18.2 & \ion{H}{2} & 9.9 & $<$0.21 & 3.0 & $<$0.76 & 0.71 & 2 \\
NGC~3256 & 10 27 51.3 & --43 54 13 & 40.1 & \ion{H}{2} & 780 & $<$1.9 & 119 & $<$7.9 & $<$20 & 2 \\
NGC~3265 & 10 31 06.8 & +28 47 48 & 23.2 & \ion{H}{2} & 30 & $<$0.78 & 5.4 & \nodata & $<$0.71 & 5 \\
NGC~3281 $^c$ & 10 31 52.1 & --34 51 13 & 49.6 & Sy2 & 17 & 47 & 58 & 46 & 173 & 2 \\
Mrk~33 & 10 32 31.9 & +54 24 03 & 22.9 & \ion{H}{2} & 59 & $<$1.0 & 45 & \nodata & 0.89 & 5 \\
3C~244.1 & 10 33 34.0 & +58 14 35 & 2353 & Sy2 & 1.4 & 0.60 & 0.30 & \nodata & \nodata & 6 \\
NGC~3310 & 10 38 45.9 & +53 30 12 & 19.8 & \ion{H}{2} & 28 & $<$0.25 & 28 & $<$1.0 & 4.0 & 8 \\
IRAS~10378+1109 & 10 40 29.2 & +10 53 18 & 641 & LINER & 3.5 & $<$1.1 & 0.60 & $<$2.4 & 2.0 & 3 \\
NGC~3351 & 10 43 57.7 & +11 42 13 & 9.30 & \ion{H}{2} & 219 & $<$2.3 & 18 & \nodata & 2.5 & 5 \\
NGC~3368 & 10 46 45.7 & +11 49 11 & 10.5 & LINER & 4.3 & $<$0.29 & 2.8 & $<$0.74 & 0.99 & 7 \\
NGC~3393 & 10 48 23.5 & --25 09 43 & 56.2 & Sy2 & \nodata & 42 & 95 & \nodata & \nodata & 6 \\
IRAS~F10565+2448 & 10 59 18.1 & +24 32 34 & 190 & \ion{H}{2} & 64 & $<$1.00 & 7.6 & $<$0.90 & $<$1.2 & 3 \\
NGC~3511 $^d$ & 11 03 23.8 & --23 05 12 & 15.9 & Sy1 & 8.7 & $<$0.49 & 1.00 & $<$1.7 & $<$1.8 & 4 \\
NGC~3507 & 11 03 25.4 & +18 08 07 & 14.2 & LINER & 3.7 & $<$0.12 & 1.4 & $<$0.55 & $<$0.21 & 2 \\
3C~249.1 & 11 04 13.7 & +76 58 58 & 1619 & Sy1 & \nodata & 0.090 & 0.14 & \nodata & $<$0.23 & 6 \\
NGC~3521 & 11 05 48.6 & --00 02 09 & 10.1 & \ion{H}{2} & 14 & $<$0.95 & 7.4 & \nodata & 2.2 & 5 \\
NGC~3516 $^c$ $^d$ & 11 06 47.5 & +72 34 06 & 38.0 & Sy1 & 7.2 & 7.3 & 16 & 9.5 & 44 & 2 \\
ESO~265-G007 & 11 07 49.6 & --46 31 27 & 11.7 & \nodata & 2.6 & $<$0.41 & 1.1 & $<$1.2 & $<$1.0 & 7 \\
NGC~3556 & 11 11 31.0 & +55 40 26 & 13.9 & \ion{H}{2} & 21 & $<$0.37 & 3.2 & $<$1.2 & $<$1.5 & 8 \\
IRAS~F11095--0238 & 11 12 03.4 & --02 54 22 & 495 & LINER & 6.1 & $<$0.48 & 1.9 & $<$1.8 & $<$0.90 & 3 \\
IRAS~F11119+3257 & 11 14 38.9 & +32 41 33 & 920 & Sy1 & 3.0 & 0.75 & 2.0 & $<$0.90 & 1.9 & 3 \\
NGC~3621 & 11 18 16.5 & --32 48 50 & 6.60 & \ion{H}{2} & 16 & 1.0 & 5.7 & \nodata & 4.6 & 5 \\
NGC~3627 & 11 20 15.0 & +12 59 29 & 9.40 & Sy2 & 23 & $<$0.92 & 8.5 & \nodata & 1.7 & 5 \\
NGC~3628 & 11 20 17.0 & +13 35 22 & 10.00 & LINER & 126 & 0.90 & 10 & $<$1.0 & $<$2.4 & 8 \\
MCG~+00-29-023 $^d$ & 11 21 12.3 & --02 59 03 & 109 & Sy2 & 47 & $<$1.0 & 4.4 & $<$5.4 & $<$6.0 & 4 \\
PG~1119+120 & 11 21 47.1 & +11 44 18 & 222 & QSO & 0.46 & 1.7 & 2.7 & 1.3 & 5.9 & 1 \\
NGC~3642 & 11 22 17.9 & +59 04 28 & 22.3 & LINER & 2.8 & $<$0.11 & 0.97 & $<$0.54 & $<$0.18 & 2 \\
NGC~3660 $^d$ & 11 23 32.3 & --08 39 30 & 53.1 & Sy2 & 6.5 & 0.98 & 1.5 & 1.7 & 3.6 & 4 \\
IRAS~11215--2806 $^c$ & 11 24 02.8 & --28 23 15 & 59.5 & Sy2 & 1.5 & 2.6 & 5.2 & 2.8 & 12 & 2 \\
NGC~3675 & 11 26 08.6 & +43 35 09 & 12.7 & \ion{H}{2} & 8.1 & $<$0.23 & 2.2 & $<$0.48 & $<$0.58 & 7 \\
IC~694 & 11 28 27.3 & +58 34 42 & 44.5 & \ion{H}{2} & 294 & $<$1.6 & 69 & $<$4.7 & $<$19 & 2 \\
NGC~3690 & 11 28 32.2 & +58 33 44 & 44.0 & Sy2 & 151 & $<$1.5 & 77 & $<$4.6 & 21 & 2 \\
PG~1126-041 & 11 29 16.7 & --04 24 07 & 269 & QSO & 1.4 & 4.3 & 5.2 & 4.7 & 16 & 1 \\
NGC~3726 & 11 33 21.1 & +47 01 45 & 14.5 & \ion{H}{2} & 5.5 & $<$0.41 & 0.43 & $<$0.52 & $<$0.44 & 7 \\
NGC~3735 $^c$ $^d$ & 11 35 57.3 & +70 32 08 & 38.9 & Sy2 & 9.4 & 7.1 & 13 & 9.8 & 37 & 2 \\
NGC~3773 & 11 38 12.9 & +12 06 43 & 11.9 & \ion{H}{2} & 18 & $<$0.81 & 16 & \nodata & 0.36 & 5 \\
NGC~3783 $^c$ & 11 39 01.7 & --37 44 18 & 41.6 & Sy1 & 18 & 16 & 27 & 12 & 40 & 2 \\
NGC~3884 & 11 46 12.2 & +20 23 29 & 101 & LINER & 1.5 & $<$0.12 & 0.86 & $<$0.72 & $<$0.72 & 2 \\
NGC~3898 & 11 49 15.4 & +56 05 03 & 16.1 & \ion{H}{2} & 1.0 & $<$0.13 & 1.3 & $<$0.51 & 0.40 & 2 \\
NGC~3938 & 11 52 49.5 & +44 07 14 & 13.3 & \ion{H}{2} & 5.4 & $<$0.66 & 1.0 & \nodata & 0.25 & 5 \\
ESO~320-G030 & 11 53 11.7 & --39 07 48 & 43.6 & \ion{H}{2} & 110 & $<$0.53 & 11 & $<$1.6 & $<$0.88 & 2 \\
NGC~3949 & 11 53 41.7 & +47 51 31 & 13.6 & \ion{H}{2} & 5.8 & $<$0.11 & 1.1 & $<$0.11 & 1.5 & 7 \\
NGC~3976 $^c$ & 11 55 57.6 & +06 45 02 & 36.0 & Sy2 & 2.3 & $<$0.26 & 1.1 & $<$0.35 & 0.73 & 2 \\
NGC~3982 $^c$ $^d$ & 11 56 28.1 & +55 07 30 & 15.2 & Sy2 & 9.2 & 3.0 & 6.7 & 1.7 & 4.6 & 2 \\
NGC~3998 & 11 57 56.1 & +55 27 12 & 14.6 & LINER & 11 & $<$0.18 & 6.9 & $<$0.65 & 0.74 & 2 \\
NGC~4013 & 11 58 31.4 & +43 56 47 & 13.8 & \ion{H}{2} & 13 & $<$0.090 & 2.8 & $<$0.89 & $<$1.0 & 7 \\
NGC~4036 & 12 01 26.8 & +61 53 44 & 19.1 & LINER & 4.4 & $<$0.060 & 2.8 & $<$0.37 & 1.4 & 2 \\
NGC~4051 $^c$ $^d$ & 12 03 09.6 & +44 31 52 & 9.27 & Sy1 & 17 & 11 & 16 & 10 & 32 & 2 \\
IRAS~F12018+1941 & 12 04 24.5 & +19 25 09 & 813 & LINER & 3.0 & $<$0.80 & 0.35 & $<$0.10 & $<$0.63 & 3 \\
NGC~4085 & 12 05 22.7 & +50 21 10 & 14.6 & \ion{H}{2} & 23 & $<$0.15 & 2.9 & $<$0.92 & $<$0.44 & 7 \\
NGC~4088 & 12 05 34.2 & +50 32 20 & 13.4 & \ion{H}{2} & 37 & $<$0.39 & 2.5 & $<$0.50 & 0.73 & 8 \\
NGC~4125 & 12 08 06.0 & +65 10 26 & 22.9 & LINER & 2.3 & $<$0.62 & 3.3 & \nodata & 0.75 & 5 \\
IRAS~12071--0444 & 12 09 45.1 & --05 01 13 & 600 & Sy2 & 5.2 & 2.9 & 5.1 & 3.7 & 6.6 & 3 \\
NGC~4151 $^c$ $^d$ & 12 10 32.6 & +39 24 20 & 14.0 & Sy1 & 131 & 77 & 205 & 68 & 244 & 2 \\
NGC~4157 & 12 11 04.4 & +50 29 04 & 13.3 & \ion{H}{2} & 12 & $<$0.070 & 1.4 & $<$0.70 & 1.1 & 7 \\
NGC~4192 & 12 13 48.3 & +14 54 01 & 19.3 & \ion{H}{2} & 19 & $<$0.18 & 4.7 & $<$0.48 & 1.6 & 2 \\
NGC~4194 & 12 14 09.5 & +54 31 36 & 40.3 & Sy2 & 165 & 3.0 & 54 & 4.0 & 27 & 8 \\
PG~1211+143 & 12 14 17.7 & +14 03 12 & 368 & QSO & 0.32 & $<$0.43 & 0.73 & $<$0.56 & 2.4 & 1 \\
NGC~4236 & 12 16 42.1 & +69 27 45 & 4.36 & \nodata & \nodata & $<$1.5 & \nodata & \nodata & $<$0.23 & 5 \\
NGC~4235 $^c$ & 12 17 09.9 & +07 11 29 & 33.5 & Sy1 & 3.8 & $<$0.43 & 3.3 & $<$0.54 & 3.3 & 2 \\
Mrk~766 $^d$ & 12 18 26.5 & +29 48 46 & 55.9 & Sy1 & 23 & 21 & 24 & 18 & 46 & 4 \\
NGC~4254 & 12 18 49.6 & +14 24 59 & 16.6 & \ion{H}{2} & 53 & $<$0.92 & 6.1 & \nodata & 2.5 & 5 \\
NGC~4258 $^c$ & 12 18 57.5 & +47 18 14 & 6.60 & Sy1 & 12 & $<$1.3 & 8.1 & $<$0.82 & 7.5 & 2 \\
NGC~4278 & 12 20 06.8 & +29 16 50 & 9.17 & LINER & 5.4 & $<$0.23 & 3.9 & $<$0.29 & 0.88 & 2 \\
NGC~4321 & 12 22 54.9 & +15 49 20 & 14.3 & LINER & 152 & $<$1.2 & 18 & \nodata & $<$1.9 & 5 \\
Mrk~52 & 12 25 42.8 & +00 34 21 & 30.1 & \ion{H}{2} & 29 & $<$0.57 & 3.8 & $<$0.83 & 1.3 & 8 \\
NGC~4388 $^c$ $^d$ & 12 25 46.7 & +12 39 43 & 36.3 & Sy2 & 75 & 45 & 106 & 68 & 308 & 2 \\
NGC~4395 $^c$ & 12 25 48.9 & +33 32 48 & 4.33 & Sy1 & 4.9 & 0.93 & 6.8 & 1.4 & 6.9 & 2 \\
NGC~4419 & 12 26 56.4 & +15 02 50 & 17.2 & \ion{H}{2} & 36 & $<$0.46 & 8.5 & $<$1.0 & $<$1.6 & 2 \\
NGC~4435 & 12 27 40.5 & +13 04 44 & 11.9 & \ion{H}{2} & 6.8 & $<$0.24 & 2.1 & $<$0.67 & 1.1 & 2 \\
NGC~4450 & 12 28 29.6 & +17 05 05 & 16.6 & LINER & 3.5 & $<$0.53 & 2.1 & \nodata & 0.63 & 5 \\
NGC~4457 & 12 28 59.0 & +03 34 14 & 12.4 & LINER & 7.8 & $<$0.18 & 3.2 & $<$0.47 & 2.2 & 2 \\
3C~273 & 12 29 06.7 & +02 03 08 & 755 & Sy1 & 1.5 & 3.4 & 6.0 & 2.9 & 8.5 & 3 \\
NGC~4477 $^c$ & 12 30 02.2 & +13 38 11 & 19.3 & Sy2 & 3.2 & $<$0.33 & 2.8 & $<$3.5 & 1.3 & 2 \\
NGC~4490 & 12 30 36.4 & +41 38 37 & 10.5 & \nodata & 3.5 & $<$0.10 & 3.5 & $<$0.10 & 1.1 & 7 \\
NGC~4486 & 12 30 49.4 & +12 23 28 & 17.0 & Sy2 & 6.9 & $<$0.19 & 5.4 & $<$0.57 & 1.2 & 2 \\
NGC~4501 $^c$ $^d$ & 12 31 59.2 & +14 25 13 & 32.7 & Sy2 & 5.0 & $<$0.47 & 4.7 & $<$0.43 & 2.5 & 2 \\
PG~1229+204 & 12 32 03.6 & +20 09 29 & 283 & QSO & 0.61 & 0.91 & 1.3 & 0.99 & 2.8 & 1 \\
NGC~4536 & 12 34 27.1 & +02 11 17 & 14.4 & \ion{H}{2} & 386 & $<$1.2 & 60 & \nodata & $<$7.9 & 5 \\
NGC~4507 $^c$ & 12 35 36.6 & --39 54 33 & 50.9 & Sy2 & 30 & 12 & 29 & 8.6 & 34 & 2 \\
NGC~4552 & 12 35 39.8 & +12 33 22 & 15.9 & LINER & 1.7 & $<$0.89 & 2.9 & \nodata & $<$0.53 & 5 \\
NGC~4559 & 12 35 57.7 & +27 57 35 & 10.3 & \ion{H}{2} & 8.3 & $<$1.0 & 2.3 & \nodata & 0.12 & 5 \\
NGC~4565 $^c$ $^d$ & 12 36 20.8 & +25 59 15 & 18.1 & Sy1 & 2.6 & 0.32 & 3.3 & $<$0.82 & 4.1 & 2 \\
NGC~4569 & 12 36 49.8 & +13 09 46 & 16.6 & LINER & 36 & $<$0.93 & 16 & \nodata & 2.9 & 5 \\
NGC~4579 $^c$ $^d$ & 12 37 43.5 & +11 49 05 & 21.5 & Sy1 & 23 & $<$0.40 & 12 & $<$0.68 & 3.2 & 2 \\
NGC~4593 $^c$ $^d$ & 12 39 39.4 & --05 20 39 & 38.8 & Sy1 & 6.8 & 3.5 & 7.4 & 5.4 & 13 & 2 \\
NGC~4594 $^c$ $^d$ & 12 39 59.4 & --11 37 23 & 15.7 & Sy1 & 14 & $<$0.55 & 16 & $<$0.71 & 2.4 & 2 \\
NGC~4602 $^d$ & 12 40 36.9 & --05 07 58 & 36.5 & Sy1 & 7.6 & 0.82 & 0.63 & $<$1.2 & $<$2.3 & 4 \\
TOL~1238-364 $^c$ $^d$ & 12 40 52.9 & --36 45 21 & 47.7 & Sy2 & 42 & 9.0 & 27 & 3.5 & 17 & 2 \\
NGC~4625 & 12 41 52.7 & +41 16 26 & 9.20 & \nodata & 5.9 & $<$0.84 & 1.2 & \nodata & $<$0.26 & 5 \\
NGC~4631 & 12 42 08.0 & +32 32 29 & 8.10 & \ion{H}{2} & 133 & $<$0.52 & 30 & \nodata & 3.0 & 5 \\
NGC~4639 $^c$ & 12 42 52.4 & +13 15 26 & 14.0 & Sy1 & 0.84 & $<$0.27 & 1.1 & $<$0.28 & $<$1.2 & 2 \\
NGC~4666 & 12 45 08.6 & --00 27 42 & 21.5 & LINER & 35 & $<$0.30 & 8.3 & 1.4 & 6.3 & 2 \\
NGC~4676 & 12 46 10.1 & +30 43 55 & 94.0 & \ion{H}{2} & 28 & $<$0.25 & 4.7 & $<$0.33 & 1.4 & 8 \\
PG~1244+026 & 12 46 35.2 & +02 22 08 & 213 & QSO & 0.94 & 0.53 & 1.2 & 1.1 & 1.5 & 1 \\
NGC~4725 $^c$ $^d$ & 12 50 26.6 & +25 30 03 & 17.4 & Sy2 & 1.2 & $<$0.38 & 2.4 & $<$0.50 & 1.8 & 2 \\
NGC~4736 & 12 50 53.1 & +41 07 13 & 5.00 & LINER & 13 & $<$0.81 & 14 & \nodata & 3.3 & 5 \\
NGC~4748 $^d$ & 12 52 12.5 & --13 24 53 & 63.4 & Sy1 & 7.4 & 6.7 & 16 & 20 & 82 & 4 \\
IRAS~F12514+1027 & 12 54 00.8 & +10 11 12 & 1667 & Sy2 & 2.3 & 1.9 & 2.7 & 1.7 & 2.7 & 3 \\
Mrk~231 & 12 56 14.2 & +56 52 25 & 186 & Sy1 & 20 & $<$3.0 & 3.0 & $<$18 & $<$9.5 & 3 \\
NGC~4826 & 12 56 43.7 & +21 40 57 & 5.00 & Sy2 & 105 & $<$1.3 & 23 & \nodata & 4.3 & 5 \\
NGC~4818 & 12 56 48.9 & --08 31 31 & 9.40 & \ion{H}{2} & 185 & $<$1.2 & 14 & $<$2.0 & $<$1.6 & 8 \\
NGC~4922 $^d$ & 13 01 24.9 & +29 18 40 & 103 & LINER & 36 & 2.4 & 9.2 & $<$1.9 & 4.3 & 2 \\
MCG~--02-33-098-E & 13 02 19.7 & --15 46 03 & 68.9 & \ion{H}{2} & 32 & $<$0.58 & 12 & $<$0.61 & 1.2 & 2 \\
MCG~--02-33-098-W & 13 02 20.4 & --15 45 59 & 68.1 & \ion{H}{2} & 67 & $<$0.51 & 12 & $<$0.79 & $<$1.2 & 2 \\
NGC~4941 $^c$ $^d$ & 13 04 13.1 & --05 33 05 & 15.9 & Sy2 & 13 & 7.2 & 24 & 6.9 & 26 & 2 \\
NGC~4939 $^c$ & 13 04 14.4 & --10 20 22 & 44.8 & Sy2 & 8.1 & 12 & 24 & 17 & 65 & 2 \\
NGC~4945 $^c$ & 13 05 27.5 & --49 28 05 & 8.19 & Sy2 & 586 & 3.4 & 69 & $<$5.1 & 35 & 2 \\
PG~1302-102 & 13 05 33.0 & --10 33 19 & 1422 & QSO & 0.36 & 0.49 & 0.66 & 0.39 & 2.6 & 1 \\
UGC~8201 & 13 06 24.9 & +67 42 25 & 3.68 & \nodata & \nodata & $<$0.75 & \nodata & \nodata & $<$0.24 & 5 \\
NGC~4968 $^d$ & 13 07 06.0 & --23 40 37 & 42.6 & Sy2 & 25 & 18 & 34 & 11 & 34 & 4 \\
PG~1307+085 & 13 09 47.0 & +08 19 48 & 739 & QSO & 0.40 & 0.56 & 0.98 & 0.62 & 0.74 & 1 \\
NGC~5005 $^c$ $^d$ & 13 10 56.2 & +37 03 33 & 13.8 & Sy2 & 37 & $<$1.1 & 13 & $<$1.2 & 4.6 & 2 \\
PG~1309+355 & 13 12 17.8 & +35 15 21 & 893 & QSO & 0.51 & 0.27 & 1.3 & $<$0.30 & $<$0.49 & 1 \\
NGC~5033 $^c$ $^d$ & 13 13 27.5 & +36 35 38 & 12.4 & Sy1 & 31 & 1.2 & 14 & 2.0 & 9.2 & 2 \\
IC~860 & 13 15 03.5 & +24 37 07 & 55.8 & \ion{H}{2} & 3.6 & $<$0.14 & 0.68 & $<$0.48 & $<$0.62 & 2 \\
IRAS~13120--5453 & 13 15 06.3 & --55 09 22 & 136 & Sy2 & 150 & 1.7 & 18 & $<$20 & 6.4 & 3 \\
NGC~5055 & 13 15 49.3 & +42 01 45 & 7.80 & LINER & 22 & $<$0.95 & 9.6 & \nodata & 2.3 & 5 \\
UGC~8387 & 13 20 35.3 & +34 08 22 & 101 & \ion{H}{2} & 113 & $<$0.54 & 19 & $<$2.3 & 6.2 & 2 \\
NGC~5104 & 13 21 23.1 & +00 20 32 & 80.0 & LINER & 43 & $<$0.39 & 5.1 & $<$1.2 & 2.4 & 2 \\
MCG~--03-34-064 $^d$ & 13 22 24.5 & --16 43 42 & 71.7 & Sy1 & 56 & 63 & 119 & 38 & 115 & 4 \\
NGC~5128 $^c$ & 13 25 27.6 & --43 01 08 & 7.85 & Sy2 & 197 & 23 & 150 & 28 & 135 & 2 \\
NGC~5135 $^c$ $^d$ & 13 25 44.1 & --29 50 01 & 58.9 & Sy2 & 112 & 14 & 58 & 18 & 73 & 2 \\
NGC~5194 $^c$ $^d$ & 13 29 52.7 & +47 11 42 & 6.77 & Sy2 & 44 & 3.0 & 34 & 3.9 & 19 & 2 \\
NGC~5195 & 13 29 59.6 & +47 15 58 & 8.00 & LINER & 18 & $<$1.0 & 6.9 & \nodata & $<$1.9 & 5 \\
3C~286 & 13 31 08.3 & +30 30 32 & 5399 & Sy1 & \nodata & 0.75 & 0.69 & \nodata & \nodata & 6 \\
NGC~5218 & 13 32 10.4 & +62 46 03 & 41.1 & \ion{H}{2} & 45 & $<$0.16 & 4.9 & $<$0.57 & $<$2.3 & 2 \\
MCG~--06-30-015 $^d$ & 13 35 53.8 & --34 17 44 & 33.4 & Sy1 & 5.0 & 5.0 & 5.9 & 7.4 & 26 & 4 \\
IRAS~F13342+3932 & 13 36 24.1 & +39 17 31 & 866 & Sy1 & 5.7 & 3.5 & 5.0 & 4.2 & 10 & 3 \\
NGC~5236 & 13 37 00.9 & --29 51 55 & 3.60 & \ion{H}{2} & 503 & $<$0.61 & 29 & $<$1.2 & 5.7 & 7 \\
Mrk~266 $^d$ & 13 38 17.5 & +48 16 37 & 116 & Sy2 & 57 & 8.0 & 28 & 11 & 53 & 8 \\
NGC~5273 $^c$ & 13 42 08.3 & +35 39 15 & 15.2 & Sy1 & 2.0 & $<$1.5 & 3.4 & $<$0.88 & 4.9 & 2 \\
Mrk~273 & 13 44 42.1 & +55 53 12 & 167 & LINER & 42 & 12 & 34 & 15 & 56 & 3 \\
IRAS~F13451+1232 & 13 47 33.4 & +12 17 24 & 565 & Sy2 & 5.0 & $<$1.0 & 5.1 & $<$2.1 & 2.1 & 3 \\
IC~4329A $^d$ & 13 49 19.3 & --30 18 34 & 69.6 & Sy1 & 28 & 29 & 57 & 35 & 117 & 4 \\
PG~1351+640 & 13 53 15.8 & +63 45 45 & 402 & QSO & 1.8 & $<$0.92 & 2.7 & $<$1.00 & $<$0.90 & 1 \\
NGC~5347 $^d$ & 13 53 17.8 & +33 29 26 & 33.6 & Sy2 & 4.2 & 2.1 & 4.1 & $<$1.7 & 7.6 & 4 \\
NGC~5371 & 13 55 39.9 & +40 27 42 & 37.1 & LINER & 1.7 & $<$0.090 & 1.1 & $<$0.32 & $<$0.50 & 2 \\
Mrk~463 & 13 56 02.9 & +18 22 19 & 222 & Sy1 & 9.3 & 18 & 41 & 20 & 69 & 3 \\
NGC~5395 $^c$ & 13 58 38.0 & +37 25 28 & 50.3 & Sy2 & \nodata & $<$0.54 & \nodata & $<$0.31 & $<$0.18 & 2 \\
NGC~5398 & 14 01 21.6 & --33 03 49 & 10.8 & \nodata & \nodata & $<$0.50 & \nodata & \nodata & $<$0.31 & 5 \\
NGC~5408 & 14 03 20.9 & --41 22 39 & 4.85 & \ion{H}{2} & \nodata & $<$0.93 & 1.2 & \nodata & $<$0.51 & 5 \\
NGC~5427 $^c$ & 14 03 26.1 & --06 01 50 & 37.5 & Sy2 & 10 & 1.7 & 5.1 & 1.2 & 4.2 & 2 \\
NGC~5474 & 14 05 01.6 & +53 39 44 & 6.40 & \ion{H}{2} & \nodata & $<$0.83 & \nodata & \nodata & $<$0.31 & 5 \\
IRAS~F14070+0525 & 14 09 31.3 & +05 11 31 & 1341 & Sy2 & 1.8 & $<$0.15 & 0.58 & $<$0.60 & $<$1.5 & 3 \\
3C~295 & 14 11 20.6 & +52 12 09 & 2571 & Sy2 & 0.060 & 0.070 & 0.050 & \nodata & \nodata & 6 \\
Circinus $^c$ & 14 13 09.9 & --65 20 20 & 6.09 & Sy2 & 427 & 219 & 379 & 261 & 871 & 2 \\
NGC~5506 $^c$ $^d$ & 14 13 14.9 & --03 12 27 & 26.1 & Sy1 & 81 & 58 & 151 & 63 & 239 & 2 \\
PG~1411+442 & 14 13 48.3 & +44 00 13 & 411 & QSO & 0.36 & 0.96 & 0.92 & 0.55 & 1.5 & 1 \\
NGC~5548 $^d$ & 14 17 59.5 & +25 08 12 & 74.5 & Sy1 & 8.5 & 5.4 & 7.3 & 3.9 & 17 & 4 \\
PG~1426+015 & 14 29 06.6 & +01 17 06 & 392 & QSO & 1.3 & 1.2 & 2.3 & 0.75 & 3.4 & 1 \\
NGC~5653 & 14 30 10.4 & +31 12 55 & 50.6 & \ion{H}{2} & 70 & $<$0.55 & 5.9 & $<$0.25 & $<$1.2 & 2 \\
NGC~5643 $^c$ & 14 32 40.8 & --44 10 28 & 16.9 & Sy2 & 38 & 25 & 56 & 30 & 121 & 2 \\
Mrk~817 $^d$ & 14 36 22.1 & +58 47 39 & 138 & Sy1 & 3.8 & 1.9 & 4.6 & 3.6 & 6.5 & 4 \\
IRAS~F14348--1447 & 14 37 38.4 & --15 00 22 & 378 & LINER & 11 & $<$0.21 & 2.6 & $<$1.5 & $<$3.3 & 3 \\
NGC~5713 & 14 40 11.5 & --00 17 20 & 29.4 & \ion{H}{2} & 127 & $<$0.80 & 17 & \nodata & 2.8 & 5 \\
IRAS~F14378--3651 & 14 40 59.0 & --37 04 32 & 306 & LINER & 11 & $<$0.90 & 1.2 & $<$2.3 & $<$3.8 & 3 \\
PG~1440+356 & 14 42 07.5 & +35 26 22 & 358 & QSO & 4.1 & 1.3 & 3.9 & 1.9 & 6.3 & 1 \\
NGC~5728 $^c$ & 14 42 23.9 & --17 15 11 & 40.9 & Sy2 & 28 & 22 & 53 & 27 & 116 & 2 \\
3C~303.1 & 14 43 14.9 & +77 07 28 & 1358 & Sy2 & 0.040 & $<$0.020 & 0.070 & \nodata & 0.090 & 6 \\
NGC~5734 & 14 45 09.1 & --20 52 13 & 58.0 & \ion{H}{2} & 77 & $<$0.96 & 8.2 & $<$0.23 & 0.93 & 2 \\
NGC~5743 & 14 45 11.0 & --20 54 48 & 59.7 & \ion{H}{2} & 50 & 1.3 & 8.9 & 0.78 & 3.2 & 2 \\
PG~1448+273 & 14 51 08.8 & +27 09 26 & 292 & QSO & 0.51 & 2.7 & 3.1 & 4.1 & 10 & 1 \\
IC~4518W & 14 57 41.2 & --43 07 55 & 68.5 & Sy1 & 40 & 25 & 49 & 24 & 85 & 2 \\
IC~4518E & 14 57 44.6 & --43 07 53 & 65.4 & \ion{H}{2} & 18 & $<$0.63 & 1.7 & $<$0.26 & 0.33 & 2 \\
3C~309.1 & 14 59 07.6 & +71 40 19 & 5833 & Sy2 & \nodata & 0.30 & 0.40 & \nodata & \nodata & 6 \\
IRAS~F15001+1433 & 15 02 31.9 & +14 21 35 & 781 & Sy2 & 6.8 & 1.1 & 2.6 & 0.66 & 1.7 & 3 \\
PG~1501+106 & 15 04 01.2 & +10 26 16 & 158 & QSO & 3.6 & 8.0 & 11 & 8.0 & 24 & 1 \\
NGC~5866 & 15 06 29.5 & +55 45 47 & 15.1 & \ion{H}{2} & 7.8 & $<$0.97 & 4.9 & \nodata & 0.93 & 5 \\
IRAS~F15091--2107 $^d$ & 15 11 59.8 & --21 19 01 & 198 & Sy1 & 12 & 8.5 & 16 & 8.1 & 31 & 4 \\
CGCG~049-057 & 15 13 13.1 & +07 13 31 & 55.8 & \ion{H}{2} & 20 & $<$0.58 & 1.5 & $<$0.82 & $<$0.59 & 2 \\
NGC~5899 $^c$ $^d$ & 15 15 03.2 & +42 02 59 & 37.0 & Sy2 & 11 & 6.8 & 16 & 6.7 & 22 & 2 \\
NGC~5907 & 15 15 53.8 & +56 19 43 & 12.1 & \ion{H}{2} & 6.1 & $<$0.0100 & 1.3 & $<$0.070 & 1.6 & 7 \\
VV~705 & 15 18 06.3 & +42 44 36 & 178 & \nodata & 62 & $<$0.33 & 14 & $<$0.79 & $<$2.8 & 2 \\
IRAS~F15206+3342 & 15 22 38.0 & +33 31 35 & 580 & \ion{H}{2} & 13 & $<$0.40 & 21 & $<$1.5 & $<$2.4 & 3 \\
NGC~5929 $^d$ & 15 26 06.2 & +41 40 14 & 35.8 & Sy2 & 13 & 1.1 & 9.8 & 2.2 & 5.3 & 4 \\
IRAS~15250+3609 & 15 26 59.4 & +35 58 37 & 245 & LINER & 10 & $<$1.2 & 2.7 & $<$2.4 & $<$1.5 & 3 \\
NGC~5936 & 15 30 00.8 & +12 59 21 & 57.1 & \ion{H}{2} & 82 & $<$0.81 & 6.1 & $<$0.63 & 1.1 & 2 \\
3C~321 & 15 31 43.5 & +24 04 19 & 441 & Sy2 & \nodata & 0.70 & 0.57 & \nodata & 1.9 & 6 \\
NGC~5953 & 15 34 32.4 & +15 11 37 & 28.9 & LINER & 51 & 1.4 & 17 & 3.0 & 17 & 2 \\
Arp~220 & 15 34 57.1 & +23 30 11 & 78.2 & \ion{H}{2} & 65 & $<$2.9 & 7.8 & $<$14 & $<$21 & 3 \\
IRAS~15335--0513 & 15 36 11.7 & --05 23 52 & 119 & \ion{H}{2} & 32 & $<$0.19 & 6.1 & $<$0.95 & $<$1.4 & 2 \\
3C~323.1 & 15 47 43.5 & +20 52 16 & 1341 & Sy1 & 0.12 & $<$0.060 & 0.11 & \nodata & 0.17 & 6 \\
MCG~--02-40-004 $^d$ & 15 48 25.0 & --13 45 27 & 110 & Sy2 & 16 & 6.1 & 8.5 & 3.1 & 13 & 4 \\
IRAS~F15462--0450 & 15 48 56.8 & --04 59 33 & 460 & Sy1 & 7.4 & $<$0.30 & 1.4 & $<$1.0 & $<$3.6 & 3 \\
IRAS~FSC15480--0344 $^d$ & 15 50 41.5 & --03 53 18 & 133 & Sy2 & 5.6 & 6.1 & 9.4 & 8.9 & 35 & 4 \\
3C~330 & 16 09 36.6 & +65 56 43 & 3170 & Sy2 & \nodata & 0.40 & 0.40 & \nodata & \nodata & 6 \\
IRAS~F16090--0139 & 16 11 40.5 & --01 47 05 & 631 & LINER & 7.8 & $<$0.12 & 2.0 & $<$2.0 & $<$1.4 & 3 \\
PG~1613+658 & 16 13 57.2 & +65 43 09 & 605 & QSO & 3.9 & 1.1 & 3.3 & 0.65 & 4.9 & 1 \\
IRAS~16164--0746 & 16 19 11.8 & --07 54 02 & 102 & LINER & 47 & 1.2 & 14 & $<$1.2 & 7.2 & 2 \\
PG~1617+175 & 16 20 11.3 & +17 24 27 & 520 & QSO & 0.29 & $<$0.17 & 0.36 & 0.28 & 0.39 & 1 \\
3C~334 & 16 20 21.8 & +17 36 23 & 3212 & Sy1 & 0.66 & 1.2 & 1.6 & \nodata & \nodata & 6 \\
PG~1626+554 & 16 27 56.1 & +55 22 31 & 626 & QSO & 0.069 & $<$0.069 & 0.11 & $<$0.16 & $<$0.20 & 1 \\
3C~343 & 16 34 33.8 & +62 45 35 & 6510 & Sy1 & 0.70 & 0.83 & 1.4 & \nodata & \nodata & 6 \\
NGC~6221 $^c$ & 16 52 46.1 & --59 13 07 & 20.8 & Sy2 & 196 & $<$1.6 & 24 & $<$3.1 & 20 & 2 \\
NGC~6240 & 16 52 58.9 & +02 24 03 & 105 & LINER & 171 & 4.4 & 61 & $<$5.7 & 27 & 3 \\
NGC~6285 & 16 58 24.0 & +58 57 21 & 82.3 & \ion{H}{2} & 18 & $<$0.080 & 3.0 & $<$0.48 & 1.2 & 2 \\
NGC~6286 & 16 58 31.4 & +58 56 10 & 79.2 & LINER & 29 & $<$0.43 & 3.3 & 1.6 & $<$0.75 & 2 \\
PG~1700+518 & 17 01 24.8 & +51 49 20 & 1505 & QSO & 1.2 & $<$0.23 & 1.6 & $<$0.43 & 1.7 & 1 \\
3C~351 & 17 04 41.4 & +60 44 30 & 1988 & Sy1 & 0.100 & 0.22 & 0.21 & \nodata & \nodata & 6 \\
IRAS~17138--1017 & 17 16 35.8 & --10 20 39 & 74.6 & \ion{H}{2} & 124 & $<$0.81 & 20 & $<$1.3 & $<$3.4 & 2 \\
NGC~6300 $^c$ & 17 16 59.5 & --62 49 14 & 15.9 & Sy2 & 11 & 13 & 15 & 9.1 & 31 & 2 \\
IRAS~F17179+5444 & 17 18 54.4 & +54 41 48 & 698 & Sy2 & 4.5 & 2.2 & 2.9 & 0.82 & 2.1 & 3 \\
IRAS~17208--0014 & 17 23 22.0 & --00 17 00 & 190 & \ion{H}{2} & 41 & $<$1.00 & 8.1 & $<$3.2 & $<$2.4 & 3 \\
NGC~6500 & 17 55 59.8 & +18 20 17 & 43.1 & LINER & 4.8 & $<$0.11 & 2.6 & $<$0.49 & $<$0.35 & 2 \\
IC~4687 & 18 13 39.6 & --57 43 31 & 73.8 & \ion{H}{2} & 146 & $<$0.80 & 45 & $<$1.2 & 2.5 & 2 \\
IC~4710 & 18 28 38.0 & --66 58 56 & 9.00 & \nodata & 0.88 & $<$1.0 & 4.3 & \nodata & $<$0.30 & 5 \\
3C~381 & 18 33 46.3 & +47 27 02 & 771 & Sy1 & 0.060 & 0.070 & 0.13 & \nodata & 0.22 & 6 \\
ESO~103-G035 & 18 38 20.3 & --65 25 39 & 56.2 & Sy2 & \nodata & 18 & 41 & \nodata & \nodata & 6 \\
IC~4734 & 18 38 25.7 & --57 29 25 & 66.7 & \ion{H}{2} & 65 & $<$0.48 & 6.1 & $<$0.66 & $<$1.8 & 2 \\
NGC~6701 & 18 43 12.5 & +60 39 12 & 56.2 & \ion{H}{2} & 73 & $<$0.59 & 7.0 & $<$0.63 & 2.1 & 2 \\
NGC~6744 & 19 09 46.1 & --63 51 27 & 9.90 & LINER & 1.1 & $<$0.030 & 1.5 & $<$0.13 & $<$0.73 & 7 \\
ESO~593-IG008 & 19 14 30.9 & --21 19 07 & 218 & \nodata & 32 & $<$0.72 & 8.9 & $<$0.29 & $<$0.78 & 2 \\
ESO~141-G055 $^d$ & 19 21 14.2 & --58 40 13 & 158 & Sy1 & 2.2 & 2.2 & 5.6 & 1.6 & 7.3 & 4 \\
IRAS~F19254--7245 & 19 31 21.4 & --72 39 18 & 283 & Sy2 & 31 & $<$2.8 & 13 & $<$1.6 & 6.3 & 3 \\
IRAS~F19297--0406 & 19 32 21.3 & --03 59 56 & 392 & \ion{H}{2} & 18 & $<$0.92 & 2.5 & $<$2.2 & $<$0.90 & 3 \\
NGC~6814 $^c$ & 19 42 40.6 & --10 19 24 & 22.4 & Sy1 & 7.2 & 3.2 & 15 & 6.1 & 27 & 2 \\
NGC~6810 $^d$ & 19 43 34.4 & --58 39 20 & 29.2 & Sy2 & 103 & $<$1.1 & 13 & $<$2.3 & 2.5 & 4 \\
NGC~6860 $^d$ & 20 08 46.9 & --61 06 00 & 64.5 & Sy1 & 5.6 & 2.8 & 6.7 & 2.4 & 12 & 4 \\
IRAS~F20087--0308 & 20 11 23.9 & --02 59 50 & 490 & LINER & 14 & $<$0.75 & 1.6 & $<$1.9 & $<$1.6 & 3 \\
IRAS~F20100--4156 & 20 13 29.5 & --41 47 34 & 610 & \ion{H}{2} & 7.3 & $<$0.48 & 2.8 & $<$1.3 & $<$4.8 & 3 \\
NGC~6890 $^d$ & 20 18 18.1 & --44 48 25 & 34.8 & Sy2 & 11 & 5.8 & 6.6 & 3.8 & 10 & 4 \\
NGC~6946 & 20 34 52.3 & +60 09 14 & 6.80 & \ion{H}{2} & 430 & $<$1.5 & 39 & \nodata & $<$7.2 & 5 \\
NGC~6951 $^c$ & 20 37 14.1 & +66 06 20 & 19.8 & Sy2 & 40 & $<$1.8 & 11 & $<$2.0 & 8.2 & 2 \\
Mrk~509 $^c$ $^d$ & 20 44 09.7 & --10 43 24 & 151 & Sy1 & 11 & 5.1 & 16 & 7.4 & 28 & 2 \\
IRAS~F20414--1651 & 20 44 18.2 & --16 40 16 & 397 & LINER & 6.8 & 1.00 & 1.6 & $<$1.5 & $<$1.8 & 3 \\
IC~5063 $^d$ & 20 52 02.3 & --57 04 07 & 49.0 & Sy2 & 27 & 30 & 66 & 24 & 114 & 4 \\
IRAS~F20551--4250 & 20 58 26.8 & --42 39 00 & 190 & \ion{H}{2} & 13 & $<$0.75 & 2.8 & $<$1.5 & $<$2.0 & 3 \\
Mrk~897 $^d$ & 21 07 45.8 & +03 52 40 & 115 & Sy2 & 24 & 1.1 & 4.4 & $<$0.80 & 0.62 & 4 \\
3C~433 & 21 23 44.5 & +25 04 11 & 470 & Sy2 & 1.9 & 2.7 & 5.2 & \nodata & $<$7.9 & 6 \\
PG~2130+099 & 21 32 27.8 & +10 08 19 & 283 & QSO & 1.4 & 3.7 & 5.7 & 4.0 & 11 & 1 \\
NGC~7130 $^c$ $^d$ & 21 48 19.5 & --34 57 04 & 69.8 & Sy2 & 82 & 7.1 & 27 & 4.2 & 15 & 2 \\
NGC~7177 & 22 00 41.2 & +17 44 17 & 16.1 & \ion{H}{2} & 4.2 & $<$0.11 & 1.5 & $<$0.48 & 0.54 & 2 \\
NGC~7172 $^c$ $^d$ & 22 02 01.9 & --31 52 11 & 37.2 & Sy2 & 31 & 8.9 & 17 & 11 & 40 & 2 \\
B2~2201+31A & 22 03 15.0 & +31 45 38 & 1523 & QSO & 0.096 & 0.53 & 0.32 & $<$0.22 & 0.56 & 1 \\
IRAS~F22017+0319 $^d$ & 22 04 19.2 & +03 33 50 & 274 & Sy2 & 5.9 & 8.3 & 14 & 9.4 & 29 & 4 \\
NGC~7213 $^c$ $^d$ & 22 09 16.2 & --47 10 00 & 25.8 & Sy1 & 24 & $<$0.94 & 13 & $<$0.73 & 2.4 & 2 \\
IC~5179 & 22 16 09.1 & --36 50 37 & 48.4 & \ion{H}{2} & 113 & $<$0.44 & 11 & $<$0.81 & 2.0 & 2 \\
PG~2214+139 & 22 17 12.3 & +14 14 20 & 297 & QSO & 0.23 & 0.27 & 0.63 & $<$0.31 & 1.3 & 1 \\
NGC~7252 & 22 20 44.8 & --24 40 41 & 66.4 & \ion{H}{2} & 42 & $<$0.33 & 3.7 & $<$0.51 & 1.3 & 8 \\
3C~445 $^d$ & 22 23 49.6 & --02 06 12 & 251 & Sy1 & 2.3 & 2.0 & 6.2 & 5.8 & 23 & 4 \\
ESO~602-G025 & 22 31 25.5 & --19 02 04 & 110 & LINER & 44 & 1.6 & 7.6 & $<$2.3 & 6.8 & 2 \\
NGC~7314 $^c$ $^d$ & 22 35 46.2 & --26 03 00 & 20.6 & Sy1 & 8.3 & 17 & 24 & 22 & 67 & 2 \\
NGC~7331 & 22 37 04.1 & +34 24 56 & 14.5 & LINER & 19 & $<$0.52 & 10 & \nodata & 3.0 & 5 \\
MCG~--03-58-007 $^d$ & 22 49 37.1 & --19 16 26 & 138 & Sy2 & 8.5 & 6.6 & 9.3 & 3.9 & 8.8 & 4 \\
IRAS~F22491--1808 & 22 51 49.3 & --17 52 23 & 354 & \nodata & 5.4 & $<$0.45 & 1.9 & $<$0.90 & $<$2.4 & 3 \\
PG~2251+113 & 22 54 10.4 & +11 36 38 & 1709 & QSO & 0.17 & 0.49 & 0.80 & 0.63 & 3.1 & 1 \\
NGC~7410 $^c$ & 22 55 00.9 & --39 39 40 & 24.0 & Sy2 & 3.7 & $<$1.1 & \nodata & $<$1.9 & $<$12 & 2 \\
NGC~7469 $^c$ $^d$ & 23 03 15.6 & +08 52 26 & 69.8 & Sy1 & 250 & 15 & 45 & 13 & 31 & 2 \\
MCG~--02-58-022 & 23 04 43.5 & --08 41 08 & 213 & Sy1 & \nodata & 2.4 & 8.6 & \nodata & \nodata & 6 \\
CGCG~453-062 & 23 04 56.5 & +19 33 08 & 109 & LINER & 22 & 2.0 & 6.6 & 2.1 & 5.5 & 2 \\
NGC~7479 $^c$ $^d$ & 23 04 56.6 & +12 19 22 & 34.2 & Sy2 & 9.4 & $<$2.4 & 5.8 & $<$2.8 & 4.9 & 2 \\
NGC~7496 $^c$ $^d$ & 23 09 47.3 & --43 25 40 & 22.8 & Sy2 & 46 & $<$0.41 & 6.3 & $<$0.98 & $<$0.48 & 2 \\
IRAS~F23128--5919 & 23 15 46.8 & --59 03 15 & 199 & Sy2 & 27 & 2.6 & 20 & 3.0 & 18 & 3 \\
NGC~7552 & 23 16 10.8 & --42 35 05 & 21.0 & LINER & 834 & $<$3.6 & 70 & \nodata & $<$10 & 5 \\
NGC~7591 & 23 18 16.3 & +06 35 08 & 70.7 & LINER & 56 & $<$0.62 & 5.5 & $<$0.76 & $<$0.54 & 2 \\
NGC~7582 $^c$ $^d$ & 23 18 23.5 & --42 22 14 & 22.7 & Sy2 & 248 & 36 & 102 & 60 & 220 & 2 \\
NGC~7590 $^c$ $^d$ & 23 18 54.8 & --42 14 20 & 23.0 & Sy2 & 6.8 & $<$0.36 & 3.7 & $<$0.52 & 3.0 & 2 \\
IRAS~F23230--6926 & 23 26 03.6 & --69 10 18 & 490 & LINER & 7.4 & $<$0.75 & 2.0 & $<$1.2 & $<$1.5 & 3 \\
NGC~7674 $^d$ & 23 27 56.7 & +08 46 44 & 127 & Sy2 & 20 & 21 & 35 & 17 & 49 & 4 \\
IRAS~F23253--5415 & 23 28 06.1 & --53 58 31 & 610 & Sy2 & 5.5 & $<$0.33 & 1.9 & 1.2 & 1.2 & 3 \\
NGC~7714 & 23 36 14.1 & +02 09 18 & 38.2 & LINER & 103 & $<$1.0 & 77 & $<$1.8 & 5.5 & 8 \\
IRAS~F23365+3604 & 23 39 01.3 & +36 21 08 & 287 & LINER & 8.6 & $<$0.80 & 0.73 & $<$0.54 & $<$2.0 & 3 \\
NGC~7743 $^c$ & 23 44 21.1 & +09 56 02 & 24.3 & Sy2 & 4.1 & $<$0.28 & 4.2 & 0.95 & 2.9 & 2 \\
CGCG381-051 $^d$ & 23 48 41.7 & +02 14 23 & 134 & Sy2 & 19 & $<$0.70 & 1.4 & $<$1.2 & $<$1.3 & 4 \\
NGC~7769 & 23 51 04.0 & +20 09 01 & 60.2 & \ion{H}{2} & 13 & $<$0.66 & 3.4 & $<$0.34 & 0.41 & 2 \\
NGC~7771 & 23 51 24.9 & +20 06 42 & 61.9 & \ion{H}{2} & 108 & $<$0.71 & 8.8 & $<$0.51 & 1.2 & 2 \\
PG~2349-014 & 23 51 56.1 & --01 09 13 & 840 & QSO & 1.4 & 0.71 & 2.0 & $<$0.96 & 3.9 & 1 \\
IRAS~F23498+2423 & 23 52 26.0 & +24 40 16 & 1046 & Sy2 & 3.2 & 1.1 & 2.7 & 2.0 & 5.0 & 3 \\
NGC~7793 & 23 57 49.8 & --32 35 27 & 3.80 & \ion{H}{2} & 10 & $<$1.0 & 2.6 & \nodata & $<$0.32 & 5
\enddata

\tablecomments{Fluxes are expressed in units of 10$^{-14}$ erg cm$^{-2}$ s$^{-1}$.}
\tablenotetext{a}{Coordinates and optical spectroscopic classification from NED.}
\tablenotetext{b}{We calculated the distance from the redshift assuming a flat cosmology with $H_0 = 70$ km s$^{-1}$Mpc$^{-1}$, $\Omega_{\rm M} = 0.3$, and $\Omega_{\rm \Lambda} = 0.7$. Except for the galaxies from \citet{Dale2009}, \citet{Bernard-Salas2009} and \citet{Goulding2009} for which we used the distances adopted by these authors.}
\tablenotetext{c}{Member of the RSA sample.}
\tablenotetext{d}{Member of the 12\micron\ sample.}
\tablerefs{(1) \citet{Veilleux2009}. (2) This work. (3) \citet{Farrah07}. (4) \citet{Tommasin08, Tommasin2010}. (5) \citet{Dale2009}. (6) \citet{Gorjian2007}. (7) \citet{Goulding2009}. (8) \citet{Bernard-Salas2009}}
\end{deluxetable*}

\end{document}